\documentclass[aps,prd,reprint,groupedaddress]{revtex4-2}

\usepackage{bm}
\usepackage{color}
\usepackage{amssymb}
\usepackage{amsmath}
\usepackage{dcolumn}
\usepackage{colortbl}
\usepackage{graphicx}
\usepackage{hyperref}
\usepackage{mathrsfs}
\usepackage{multirow}
\usepackage{longtable}
\usepackage[table,xcdraw]{xcolor}

\usepackage{url}
\usepackage{ulem}
\usepackage{natbib}
\usepackage{diagbox}
\usepackage{rotating}
\usepackage{enumitem}
\usepackage{booktabs}
\usepackage{threeparttable}
\usepackage[thicklines]{cancel}

\begin{document}

\title{Impact of $\Xi$-Hypernuclear Constraints on Relativistic Equation of State and Properties of Hyperon Stars}

\author{Shi Yuan Ding}
\affiliation
{MOE Frontiers Science Center for Rare Isotopes, Lanzhou University, Lanzhou 730000, China}
\affiliation
{School of Nuclear Science and Technology, Lanzhou University, Lanzhou 730000, China}

\author{Xiang Dong Sun}
\affiliation
{Department of Astronomy, Xiamen University, Xiamen, Fujian 361005, China}

\author{Bao Yuan Sun\footnote{
		Corresponding author (Email: sunby@lzu.edu.cn)}}
\affiliation
{MOE Frontiers Science Center for Rare Isotopes, Lanzhou University, Lanzhou 730000, China}
\affiliation
{School of Nuclear Science and Technology, Lanzhou University, Lanzhou 730000, China}

\author{Ang Li\footnote{
		(Email: liang@xmu.edu.cn)}}
\affiliation
{Department of Astronomy, Xiamen University, Xiamen, Fujian 361005, China}

\begin{abstract}
Significant uncertainties persist in describing the internal structure and equation of state of hyperon stars due to the limited understanding of the mechanisms underlying hyperon interactions. Constraining the interaction parameter space through a combination of the latest astronomical observations and hypernuclear physics experiments is therefore essential. In this study, we incorporate experimental constraints from $\Xi$ hypernuclear physics on top of $\Lambda$ hyperons considered in Ref. \cite{Sun2023APJ942.55}. Specifically, based on updated measurements of hyperon separation energies from $\Xi$ hypernuclear experiments, three sets of $\Xi N$ effective interactions are constructed and a linear correlation between their scalar ($\sigma$) and vector ($\omega$) coupling strength ratios is proposed as a constraint derived from $\Xi$ hypernuclear physics. Together with experimental correlations and astronomical observational data, four types of analyses are performed to constrain hyperon-nucleon interactions and the properties of hyperon stars. Compared to $\Lambda$ hyperons, the shallower potential and larger rest mass of $\Xi$ hyperons lead to a higher threshold for their appearance in hyperon stars. As a result, the parameter space for the hyperon-nucleon interaction of $\Xi$ hyperons remains more loosely constrained. In particular, compared to the vector $\omega$ meson-hyperon coupling, the introduction of linear correlations in hypernuclear physics imposes a more substantial constraint on the scalar $\sigma$ meson-hyperon coupling, significantly enhancing its coupling strength and ensuring the stiffness of the equation of state, highlighting the crucial role of hypernuclear studies in solving the hyperon puzzle. Consequently, a maximum mass of around $2M_{\odot}$ can be achieved with all four interactions considered in this study under the combined constraints from astronomical observations and nuclear physics. Moreover, the uncertainties in both the fractions and the threshold densities at which hyperons appear inside neutron stars are notably reduced, along with those in the mass-radius predictions. The corresponding uncertainties associated with hyperon-nucleon contributions and the impact on structural properties of hypernuclei are also reliably estimated.
\end{abstract}

\maketitle

\section{Introduction}

Neutron stars (NSs) are considered as the most compact objects, with interior densities spanning a wide range and potentially reaching up to about $ 10 $ times normal nuclear density. This suggests that, besides nucleonic degrees of freedom, non-nucleonic particles may also be present \cite{Ambartsumyan1960SVA4.187, Malik2022PRD106.063024, Carvalho2024PRD110.123016}. 
In fact, as early as the 1960s, the possibility of hyperon appearance in free dense matter was explored based on thermodynamic considerations \cite{Ambartsumyan1960SVA4.187}. Research has shown that when the baryon density exceeds a certain threshold, hyperons can emerge inside NS matter. Moreover, unlike unstable hyperons that decay rapidly in vacuum, the extremely high density of the NS core, along with the Pauli blocking, leading to the existence of hyperonic matter. The corresponding NSs are often referred as hyperon stars. Although the emergence of hyperonic matter is energetically favorable, it poses a challenge. Hyperons introduce new channels for filling the Fermi sea, reducing the degeneracy pressure and softening the equation of state (EOS) of NS matter. This may significantly lower the maximum mass of NSs, potentially below $2M_{\odot}$, contradicting observations of pulsars with masses around two solar masses, known as the ``hyperon puzzle" \cite{Schulze2006PRC73.058801, Vidana2013NPA914.367, Haidenbauer2017EPJA53.121}. To resolve this puzzle, various mechanisms have been proposed \cite{Vidana2013NPA914.367}, including the introduction of repulsive hyperonic three-body forces \cite{Yamamoto2014PRC90.045805, Lonardoni2015PRL114.092301, Wirth2016PRL117.182501, Friedman2023PLB837.137669}, the inclusion of $\Delta$-resonances \cite{Sedrakian2020PRD102.041301, Sedrakian2023PPNP131.104041, Vishal2025PRD112.023016}, a possible phase transition to deconfined quark matter \cite{Weissenborn2011APJL740.L14, Bonanno2012A&A539.A16, Klahn2013PRD88.085001, Albino2024PRD110.083037} and the interaction of hyperons with vector mesons \cite{Schaffner1996PRC53.1416, Weissenborn2013NPA914.421, Tolos2017PASA34.e065, Fortin2017PRC95.065803, Biswa2019APJ885.25, Lopes2021NPA1009.122171, Tu2022APJ925.16}. Notably, understanding the "hyperon puzzle" is highly sensitive to the specific details of the interactions. However, the evolution of nuclear forces under different density conditions within NSs remains shrouded in many unresolved mysteries. To clarify the internal composition and EOS of NS cores, a deeper understanding of interactions in high-density regions is crucial. To this end, combining astronomical observations with relevant nuclear physics experimental data may provide key constraints on interactions under extreme density conditions \cite{Beznogov2023PRC107.045803, Salinas2023PRC107.045802, Zhu2023APJ943.163, Beznogov2024APJ966.216, Sun2025SSPMA}.

Hyperon-nucleon ($YN$) scattering experiments provide the most direct probe of $YN$ interactions. However, the hyperon's extremely short lifetime makes a direct measurement of its scattering cross section extraordinarily challenging, and the experimental data are correspondingly sparse. Researchers typically employ theoretical methods to extract interaction details from these limited $YN$ scattering results. A widely used strategy leverages flavor SU(3) symmetry in baryon-baryon forces to relate the $YN$ and $NN$ systems, thereby imposing constraints on the $YN$ interaction. Based on this idea, a variety of theoretical methods have been developed for constructing baryon-baryon interactions, including the Nijmegen potential models \cite{Nagels1977PRD15.2547, Rijken1999PRC59.21, Vidana2001PRC64.044301, Rijken2010Prog.Theor.Phys.Suppl.185.14, Schulze2010NPA835.19}, the J\"{u}lich multi-meson exchange models \cite{Holzenkamp1989NPA500.485, Reuber1994NPA570.543, Haidenbauer2005PRC72.044005}, the chiral effective field theory \cite{Savage1996PRD53.349, Machleidt2011Phys.Rep.503.1, Haidenbauer2013NPA915.24, Li2016PRD94.014029, Ren2020PRC101.034001, Liu2021PRC103.025201, Zheng2025PLB864.139416}, and lattice QCD \cite{Beane2007NPA794.62, Inoue2010Prog.Theor.Phys.124.591, Beane2011PPNP66.1, Aoki2012PTEP2012.01A105, Beane2012PRL109.172001, Beane2013PRD87.034506, Sasaki2015PTEP2015.113B01}. Furthermore, as hypernuclear properties are highly sensitive to the $YN$ and $YY$ interactions, information related to hyperons can also be obtained through hypernuclear studies. To date, over $40$ single-$\Lambda$ hypernuclei have been identified, along with some double-$\Lambda$ and single-$\Xi$ hypernuclei \cite{Hashimoto2006Prog.Part.Nucl.Phys.57.564, Feliciello2015Rep.Prog.Phys.78.096301, Gal2016Rev.Mod.Phys.88.035004}. As the hypernuclear system with the smallest strangeness, single-$\Lambda$ hypernuclei have yielded experimental data on various properties, such as $\Lambda$ separation energy, across different mass regions, leading to extensive theoretical research \cite{Brockmann1977PLB69.167, Bouyssy1981PLB99.305, Jennings1994PRC49.2472, Shen2002NPA707.469, Zhou2007PRC76.034312, Hiyama2009Prog.Part.Nucl.Phys.63.339, Bogner2010Prog.Part.Nucl.Phys.65.94, Hu2014PRC90.014309, Lu2014PRC89.044307, Wirth2014PRL113.192502, Gazda2016PRL116.122501, Wirth2016PRL117.182501, Xia2017Sci.China-Phys.Mech.Astron.60.102021, Tanimura2019PRC99.034324, Rong2021PRC104.054321, Zhang2021PRC103.034321, Ding2022PRC106.054311, Ding2023CPC47.124103, Xue2024PRC109.024324}. For $\Xi$ hypernuclei, early experimental evidence was ambiguous, and theoretical modeling of the interaction often relied on the assumption of a given hyperon potential (e.g., $U_{\Xi}=-14$ MeV) \cite{Khaustov2000PRC61.054603, Hiyama2008PRC78.054316, Gal2016Rev.Mod.Phys.88.035004}. Consequently, theoretical studies on $\Xi$ hypernuclear properties remained limited until the first observation of a bound $^{15}_{\Xi}$C hypernucleus in 2015 \cite{Nakazawa2015Prog.Theor.Exp.Phys.2015.033D02}, which has since spurred renewed theoretical interest \cite{Sun2016PRC94.064319, Hu2017PRC96.054304, Liu2018PRC98.024316, Hiyama2020PRL124.092501, Jin2020EPJA56.135, Guo2021PRC104.L061307, Le2021EPJA57.339, Friedman2021PLB820.136555, Hu2022JPG49.025104, Tanimura2022PRC105.044324, Friedman2023PLB837.137640, Isaka2024PRC109.044317, Ding2025PRC111.014301}. These studies have shown that, unlike the $\Lambda$, the structure of $\Xi$ hypernuclei involves additional isospin-dependent effects \cite{Jennings1994PRC49.2472, Ding2025PRC111.014301}. Thus, such research holds promise for playing a significant role in probing isospin-sensitive physics within NSs \cite{TeodorodosSantos2025PRC111.035805}.

While experimental studies and structural analyses of hypernuclei have provided valuable constraints on hyperon-related interactions, substantial uncertainties persist in applying these results to the description of hyperonic stars. To more effectively constrain the EOS and interaction properties of hyperonic matter in NSs, it is necessary to integrate hypernuclear physics with astrophysical research, forming joint constraints. Along this line, our previous work \cite{Sun2023APJ942.55} performed a Bayesian inference of the coupling strengths between the $\Lambda$ hyperon and mesons by comparing phenomenological $\Lambda$ interactions with observed NS properties . This analysis incorporated tidal measurements from the GW170817 binary NS merger observed by LIGO/Virgo \cite{Abbott2017PRL119.161101}, along with mass-radius observations of the pulsars PSR J0030+0451 \cite{Riley2019APJL887.L21, Miller2019APJL887.L24} and PSR J0740+6620 \cite{Riley2021APJL918.L27, Miller2021APJL918.L28} from NICER. The results indicate that, under astrophysical constraints alone, the scalar coupling strength ratio of the $\Lambda$ hyperon, $R_{\sigma\Lambda}$ ($R_{\phi Y}=g_{\phi Y}/g_{\phi N}$), tends to favor smaller values, while the vector coupling strength ratio, $R_{\omega\Lambda}$, is inclined toward larger values. When experimental constraints from $\Lambda$ hypernuclei are included, an enhancement of the scalar coupling is required to reconcile the strong vector coupling necessary for hyperonic star matter \cite{Sun2023APJ942.55}. 

It is noteworthy that previous studies on hypernuclear properties typically did not alter the $\Xi N$ interaction, or adjusted it only within empirical bounds, due to the lack of theoretical support for hypernuclear structure research \cite{Fortin2020PRD101.034017, Malik2022PRD106.063024, Sun2023APJ942.55, Huang2024MNRAS536.3262, Vishal2025PRD112.023016}. As another crucial component of hyperon stars, the interaction of $\Xi$ hyperons may also significantly affect the description of their properties. Recently, updated experimental data and developments in the theoretical models prompted us to undertake systematic investigations into various hypernuclei and establish a series of $YN$ interactions. The results demonstrate that these interactions exhibit significant nuclear in-medium effects as density evolves \cite{Ding2022PRC106.054311, Ding2023CPC47.124103, Yang2024PRC110.054320, Ding2025PRC111.014301}. These findings provide new hypernuclear physics constraints for integration into the Bayesian analysis of the present work. As part of this series of related investigations, this work employs the methodology of Ref. \cite{Sun2023APJ942.55}. All calculations are performed within the relativistic mean-field (RMF) framework, with particular focus on analyzing the effects of introducing $\Xi$ hypernuclear physics constraints on the $YN$ interactions and the resulting properties of hyperon stars.

The paper is organized as follows: In Section \ref{sec:theoretical}, we briefly introduce the RMF model, extract the $\Xi N$ interaction from existing $\Xi$ hypernuclear data, establish a linear correlation between the scalar coupling strength ratio $R_{\sigma\Xi}$ and vector coupling strength ratio $R_{\omega\Xi}$ of the $\Xi$ hyperon as a likelihood for Bayesian inference, and discuss NSs composed purely of nucleonic matter based on selected RMF parameter sets. The NS observations used, the empirical relation between scalar and vector coupling strength ratios of hyperons derived from hypernuclear experimental data, and the Bayesian analysis method are described in Section \ref{sec:analysis}. We present our results and a discussion in Section \ref{sec:result} and summarize our paper in Section \ref{sec:summary}.

\section{Relativistic EOS for Hypernuclear Matter} \label{sec:theoretical}

\subsection{The RMF model}

The relativistic mean field theory has been widely applied to the study of finite nuclei and nuclear matter, owing to its ability to provide a self-consistent description of almost all nuclei across the nuclear chart \cite{Reinhard1989Rep.Prog.Phys.52.439, Ring1996Prog.Part.Nucl.Phys.37.193, Vretenar2005Phys.Rep.409.101, Meng2006Prog.Part.Nucl.Phys.57.470, Niksic2011Prog.Part.Nucl.Phys.66.519, Meng2015JPG42.093101, Meng2016Relativistic-Density-Functional-for-Nuclear-Structure}. With the advancement of theoretical models, RMF has further been extended to describe hypernuclear systems involving strangeness degrees of freedom \cite{Brockmann1977PLB69.167, Glendenning1991PRL67.2414, Jennings1994PRC49.2472, Vretenar1998PRC57.R1060, Win2008PRC78.054311, Lu2011PRC84.014328, Tanimura2012PRC85.014306, Lu2014PRC89.044307, Sun2016PRC94.064319, Wu2017PRC95.034309, Xia2017Sci.China-Phys.Mech.Astron.60.102021, Liu2018PRC98.024316, Tanimura2019PRC99.034324, Rong2020PLB807.135533, Chen2021Sci.ChinaPhys.Mech.Astron.64.282011, Rong2021PRC104.054321, Ding2022PRC106.054311, Ding2023CPC47.124103, Xia2023Sci.China-Phys.Mech.Astron66.252011, Yao2024NPA1042.122794, Yang2024PRC110.054320, Ding2025PRC111.014301}. As emphasized in Ref. \cite{Sun2023APJ942.55}, the covariant nature of this theoretical framework enables it to, to some extent, avoid the additional uncertainty in the EOS caused by hyperon-related three-body interactions. Since this study aims to explore the potential effects of the additional introduced $\Xi$ hypernuclear physics constraints on the properties of hyperonic matter, we follow the selection strategy for RMF effective interactions proposed in Ref. \cite{Ding2025PRC111.014301}. We employ density-dependent (DD) RMF effective interactions, namely TW99 \cite{Typer1999NPA656.331}, PKDD \cite{Long2004PRC69.034319}, DD-ME2 \cite{Lalazissis2005PRC71.024312}, DD-MEX \cite{Taninah2020PLB800.135065, Rather2021PRC103.055814}, DD-ME$\delta$ \cite{Roca-Maza2011PRC84.054309}, and DD-LZ1 \cite{Wei2020CPC44.074107} to account for the in-medium effects in the nuclear force. Based on these effective Lagrangians, a series of $\Xi N$ interactions are constructed using the experimental separation energies of $\Xi$ hypernuclei, enabling us to explore interplay of different hyperon interaction channels.

To describe hypernuclear matter within the framework of RMF theory, the covariant Lagrangian density provides the foundation, which is
\begin{align}
	\mathscr{L} = \mathscr{L}_{B} + \mathscr{L}_{\varphi} + \mathscr{L}_{\rm{int}} + \mathscr{L}_{l},
\end{align}
The free baryonic Lagrangian is given by the sum of the Dirac Lagrangians for individual baryons with mass $M_{B}$
\begin{align}
	\mathscr{L}_{B} = &\sum_{B}\bar{\psi}_{B}\left(i\gamma^{\mu}\partial_{\mu}-M_{B} \right)\psi_{B},
\end{align}
where the index $B$ sums over the baryonic octet ($n, p, \Lambda, \Sigma^{-}, \Sigma^{0}, \Sigma^{+}, \Xi^{-}, \Xi^{0}$), and $\psi_B$ denotes the Dirac field of baryon $B$. The Lagrangian for the free meson and photon fields is given by
\begin{align}
	\mathscr{L}_{\varphi} =
	&+\frac{1}{2}\partial^{\mu}\sigma\partial_{\mu}\sigma-\frac{1}{2}m_{\sigma}^{2}\sigma^{2}-\frac{1}{4}\Omega^{\mu\nu}\Omega_{\mu\nu}+\frac{1}{2}m_{\omega}^2\omega^{\mu}\omega_{\mu}\notag\\
	&+\frac{1}{2}\partial^{\mu}\vec{\delta}\cdot\partial_{\mu}\vec{\delta}-\frac{1}{2}m_{\delta}^{2}\vec{\delta}^{2}-\frac{1}{4}\vec{R}^{\mu\nu}\cdot\vec{R}_{\mu\nu}+\frac{1}{2}m_{\rho}^2\vec{\rho}^{\mu}\cdot\vec{\rho}_{\mu}\notag\\
	&-\frac{1}{4}F^{\mu\nu}F_{\mu\nu},
\end{align}
$\Omega^{\mu\nu}$, $\vec{R}^{\mu\nu}$, and $F^{\mu\nu}$ are the field tensors for the vector mesons $\omega^\mu$ and $\vec{\rho}^\mu$, and for the photon $A^\mu$, respectively. $m_\phi$ ($\phi = \sigma, \omega^\mu, \vec{\delta}, \vec{\rho}^\mu$) denotes the mass of the corresponding meson. The interaction between baryons and mesons (photons) is described by the Lagrangian $\mathscr{L}_{\rm{int}}$.
\begin{align}
	\mathscr{L}_{\rm{int}} = &\sum_{B}\bar{\psi}_{B}\left(-g_{\sigma B}\sigma-g_{\omega B}\gamma^{\mu}\omega_{\mu} \right. \notag\\
	&\left. -g_{\delta B}\vec{\tau}_{B}\cdot\vec{\delta} -g_{\rho B}\gamma^{\mu}\vec{\tau}_{B}\cdot\vec{\rho}_{\mu}-e\gamma^{\mu}Q_{B}A_{\mu}\right)\psi_{B} \notag\\
	&-\bar{\psi}_{Y}\frac{f_{\omega Y}}{2M_{Y}}\sigma^{\mu\nu}\partial_{\nu}\omega_{\mu}\psi_{Y},
\end{align}
Here, $Y = \Lambda, \Sigma^{-}, \Sigma^{0}, \Sigma^{+}, \Xi^{-}, \Xi^{0}$. The term $\frac{f_{\omega Y}}{2M_{Y}}$ denotes the tensor coupling between hyperons and the $\omega$ meson field. The baryon charge number $Q_{B}$ and the third component of the isospin operator, $\tau_{3,B}$, can be expressed as follows:
\begin{align}
	Q_{B} = 
	\begin{cases}
		0, & \text{for } \Lambda, \\[4pt]
		\dfrac{1 - \tau_{3,N}}{2}, & \text{for } n,\ p, \\[8pt]
		-\dfrac{1 + \tau_{3,\Xi}}{2}, & \text{for } \Xi^0,\ \Xi^-, \\[8pt]
		-\tau_{3,\Sigma}, & \text{for } \Sigma^+,\ \Sigma^0,\ \Sigma^-.
	\end{cases}
\end{align}

\begin{align}
	\tau_{3,B} = 
	\begin{cases} 
		0, & B = \Lambda, \\[4pt]
		+1,\ -1, & B = n,\ p, \\[4pt]
		-1,\ +1, & B = \Xi^0,\ \Xi^-, \\[4pt]
		-1,\ 0,\ +1, & B = \Sigma^+,\ \Sigma^0,\ \Sigma^-.
	\end{cases}
\end{align}

For hyperonic matter, the effective Lagrangian includes lepton contributions
\begin{align}
	\mathcal{L}_{l} = &\sum_{l}\bar{\psi}_{l}\left(i\gamma^{\mu}\partial_{\mu}-M_{l} \right)\psi_{l},
\end{align}
where $\psi_{l}$ denotes the lepton fields (with $l$ summing over electrons and muons) and $m_{l}$ is the corresponding lepton mass. At the mean-field level, the many-body state is constructed as a Slater determinant of single-particle wave functions, described by four-component Dirac spinors. The Klein-Gordon equations for the meson fields and the Dirac equations for the baryon fields are solved self-consistently in the RMF approximation, with the meson field operators are replaced by their expectation values in the ground state.

\subsection{$\Xi N$ effective interaction in RMF models}

\begin{table*}[hbpt]
	\centering
	\caption{A series of $\Xi N$ effective interactions is constructed for each RMF effective Lagrangian by varying the $\omega$–$\Xi$ coupling strength ratio $R_{\omega\Xi}$ from $0.200$ to $0.900$. For each chosen value of $R_{\omega\Xi}$, the $\sigma$–$\Xi$ coupling strength ratio $R_{\sigma\Xi}$ is determined by fitting to experimental data on the $\Xi^{-}$ separation energies, namely the $1s$ state of $^{15}_{\Xi^{-}}$C \cite{Yoshimoto2021PTEP2021.073D02} ($\Xi$C$s$), the $1p$ state of $^{15}_{\Xi^{-}}$C \cite{Hayakawa2021PRL126.062501} ($\Xi$C$p$), and the $1p$ state of $^{13}_{\Xi^{-}}$B \cite{Aoki2009NPA828.191} ($\Xi$B$p$), see the text for details. The coupling strengths for other meson–hyperon channels are fixed at $R_{\rho\Xi}=1.000$ and $R_{\delta\Xi}=1.000$, with the additional constraint that the $\omega$–$\Xi$ tensor coupling is set to $f_{\omega\Xi}=-0.400g_{\omega\Xi}$.}\label{Tab:CouplingStrengthX}
	\renewcommand{\arraystretch}{1.5}
	\doublerulesep 0.1pt \tabcolsep 16pt
	\begin{tabular}{cccccccc}
		\hline
		& $R_{\omega\Xi}$ & \multicolumn{6}{c}{$R_{\sigma\Xi}$}                                  \\ \cline{3-8} 
		&                 & TW99     & PKDD     & DD-ME2   & DD-MEX   & DD-ME$\delta$ & DD-LZ1   \\ \hline
		\multirow{9}{*}{$\Xi$C$s$} & 0.200           & 0.200130 & 0.204435 & 0.205060 & 0.200446 & 0.214409      & 0.195732 \\
		& 0.300           & 0.282081 & 0.285830 & 0.286412 & 0.282596 & 0.293437      & 0.278225 \\
		& 0.333           & 0.309146 & 0.312701 & 0.313264 & 0.309712 & 0.319534      & 0.305429 \\
		& 0.400           & 0.364122 & 0.367271 & 0.367784 & 0.364769 & 0.372539      & 0.360609 \\
		& 0.500           & 0.446235 & 0.448742 & 0.449154 & 0.446939 & 0.451703      & 0.442780 \\
		& 0.600           & 0.528407 & 0.530226 & 0.530503 & 0.529082 & 0.530919      & 0.524636 \\
		& 0.700           & 0.610619 & 0.611709 & 0.611815 & 0.611173 & 0.610176      & 0.606093 \\
		& 0.800           & 0.692856 & 0.693178 & 0.693075 & 0.693194 & 0.689465      & 0.687095 \\
		& 0.900           & 0.775104 & 0.774624 & 0.774274 & 0.775130 & 0.768779      & 0.767608 \\ \hline
		\multirow{9}{*}{$\Xi$C$p$} & 0.200           & 0.210574 & 0.213892 & 0.215572 & 0.212931 & 0.220320      & 0.217157 \\
		& 0.300           & 0.292094 & 0.294496 & 0.295747 & 0.293879 & 0.298810      & 0.296555 \\
		& 0.333           & 0.318984 & 0.321078 & 0.322175 & 0.320552 & 0.324709      & 0.322607 \\
		& 0.400           & 0.373558 & 0.375021 & 0.375789 & 0.374644 & 0.377284      & 0.375274 \\
		& 0.500           & 0.454958 & 0.455461 & 0.455696 & 0.455216 & 0.455738      & 0.453313 \\
		& 0.600           & 0.536288 & 0.535815 & 0.535467 & 0.535594 & 0.534173      & 0.530690 \\
		& 0.700           & 0.617547 & 0.616083 & 0.615105 & 0.615778 & 0.612588      & 0.607442 \\
		& 0.800           & 0.698734 & 0.696268 & 0.694617 & 0.695779 & 0.690986      & 0.683617 \\
		& 0.900           & 0.779851 & 0.776377 & 0.774013 & 0.775607 & 0.769369      & 0.759273 \\ \hline
		\multirow{9}{*}{$\Xi$B$p$} & 0.200           & 0.217303 & 0.220715 & 0.221986 & 0.218690 & 0.228127      & 0.220715 \\
		& 0.300           & 0.299074 & 0.301654 & 0.302557 & 0.300116 & 0.306807      & 0.301371 \\
		& 0.333           & 0.326115 & 0.328357 & 0.329128 & 0.326959 & 0.332777      & 0.327859 \\
		& 0.400           & 0.380911 & 0.382561 & 0.383043 & 0.381413 & 0.385513      & 0.381430 \\
		& 0.500           & 0.462699 & 0.463429 & 0.463436 & 0.462568 & 0.464240      & 0.460855 \\
		& 0.600           & 0.544465 & 0.544255 & 0.543734 & 0.543575 & 0.542987      & 0.539633 \\
		& 0.700           & 0.626206 & 0.624908 & 0.623937 & 0.624430 & 0.621753      & 0.617777 \\
		& 0.800           & 0.707920 & 0.705771 & 0.704049 & 0.705136 & 0.700536      & 0.695315 \\
		& 0.900           & 0.789604 & 0.786463 & 0.784075 & 0.785700 & 0.779339      & 0.772295 \\ \hline
	\end{tabular}
\end{table*}

Before incorporating the physical constraints of the $\Xi$ hypernucleus into Bayesian inference, this section details the construction of the $\Xi N$ interaction. Within the meson-exchange picture of the RMF framework, the $\Xi N$ interaction relates to the coupling strengths among the mesons and $\Xi$ hyperon involved in the interaction. The effective Lagrangian employed in this study is consistent with that in Ref. \cite{Ding2025PRC111.014301}, with meson-hyperon coupling strengths determined by symmetry considerations and the latest $\Xi$ hypernuclei experimental data. Specifically, the isovector-vector coupling ratio is fixed as $R_{\rho\Xi}=g_{\rho\Xi}/g_{\rho N}=1.000$ in accordance with the SU(3) Clebsch-Gordan coefficients \cite{Jennings1994PRC49.2472}. The isovector-scalar coupling ratio is similarly set to $R_{\delta\Xi}=g_{\delta\Xi}/g_{\delta N}=1.000$ \cite{Shao2010PRC82.055801, Tu2022APJ925.16}. A tensor coupling is also included in the hyperon channel, with $f_{\omega\Xi}=-0.4g_{\omega\Xi}$. To explore the possible correlation between the isoscalar-scalar and isoscalar-vector couplings of the $\Xi$ hyperon, we relax the SU(3) symmetry constraint and allow $R_{\omega\Xi}$ to vary over a range from $0.200$ to $0.900$. For each value of $R_{\omega\Xi}$, the corresponding isoscalar-scalar coupling ratio $R_{\sigma\Xi}$ is determined by reproducing the experimental $\Xi^{-}$ separation energy $B_{\Xi^{-}}$, defined as:
\begin{align}
	B_{\Xi^{-}}[A]&\equiv E[n,p,-]-E[n,p,\Xi^{-}]\notag\\
	&=E[^{A-1}(Z+1)]-E[^{A}_{\Xi^{-}}Z],
\end{align}
where $E$ denotes the binding energy of either the hypernucleus or its nucleonic core, and $A=Z+N+1$ for the hypernucleus. Although the meson-baryon coupling strengths in DDRMF models are density-dependent, the ratios $R_{\phi\Xi}$ remain constant. The mass of the $\Xi^{-}$ hyperon is taken as $M_{\Xi^{-}}=1321.7$ MeV.

As emphasized in our previous study~\cite{Ding2025PRC111.014301}, the uncertainty in experimental data on the $\Xi^{-}$ hyperon separation energy makes the choice of an appropriate fitting target essential for constructing the $\Xi N$ interaction and reliably describing hypernuclear structure. The deeply bound hypernucleus $^{15}_{\Xi^{-}}$C, first conclusively identified in experiments with an attractive $\Xi N$ interaction, is an ideal candidate for this purpose. In addition, the $^{13}_{\Xi^{-}}$B system is also frequently used to constrain the interaction, as its study in relevant experiments avoids the uncertainties associated with the $\Xi^{-}$-$\Xi^{0}$ mixing. Therefore, based on selected DDRMF effective interactions, including the TW99, PKDD, DD-ME2, DD-MEX, DD-ME$\delta$, and DD-LZ1, we construct the $\Xi N$ effective interaction by adjusting the coupling ratio $R_{\omega\Xi}$ and determining $R_{\sigma\Xi}$ through three fitting strategies.

The first strategy adopts the separation energy of the $1s$ state of the $\Xi^{-}$ hyperon in $^{15}_{\Xi^{-}}$C as a fitting constraint. Experimental information on this state is provided by the IRRAWADDY and KINKA events~\cite{Yoshimoto2021PTEP2021.073D02}. However, the significant discrepancy between the separation energies reported in these two events poses a challenge for setting reliable constraints on theoretical models. To reduce this uncertainty, Ref.~\cite{Yoshimoto2021PTEP2021.073D02} proposed a weighted average of the IRRAWADDY value $B_{\Xi^{-}} = 6.27 \pm 0.27$ MeV and the larger KINKA value $B_{\Xi^{-}} = 8.00 \pm 0.77$ MeV, yielding a reference of $B_{\Xi^{-}} = 6.46 \pm 0.25$ MeV. In our work, the central value of this reference, $B_{\Xi^{-}} = 6.46$ MeV, is adopted as the fitting target to determine $R_{\sigma\Xi}$, with the corresponding interaction denoted as $\Xi$C$s$.
\begin{figure}[htbp]
	\centering
	\includegraphics[width=0.48\textwidth]{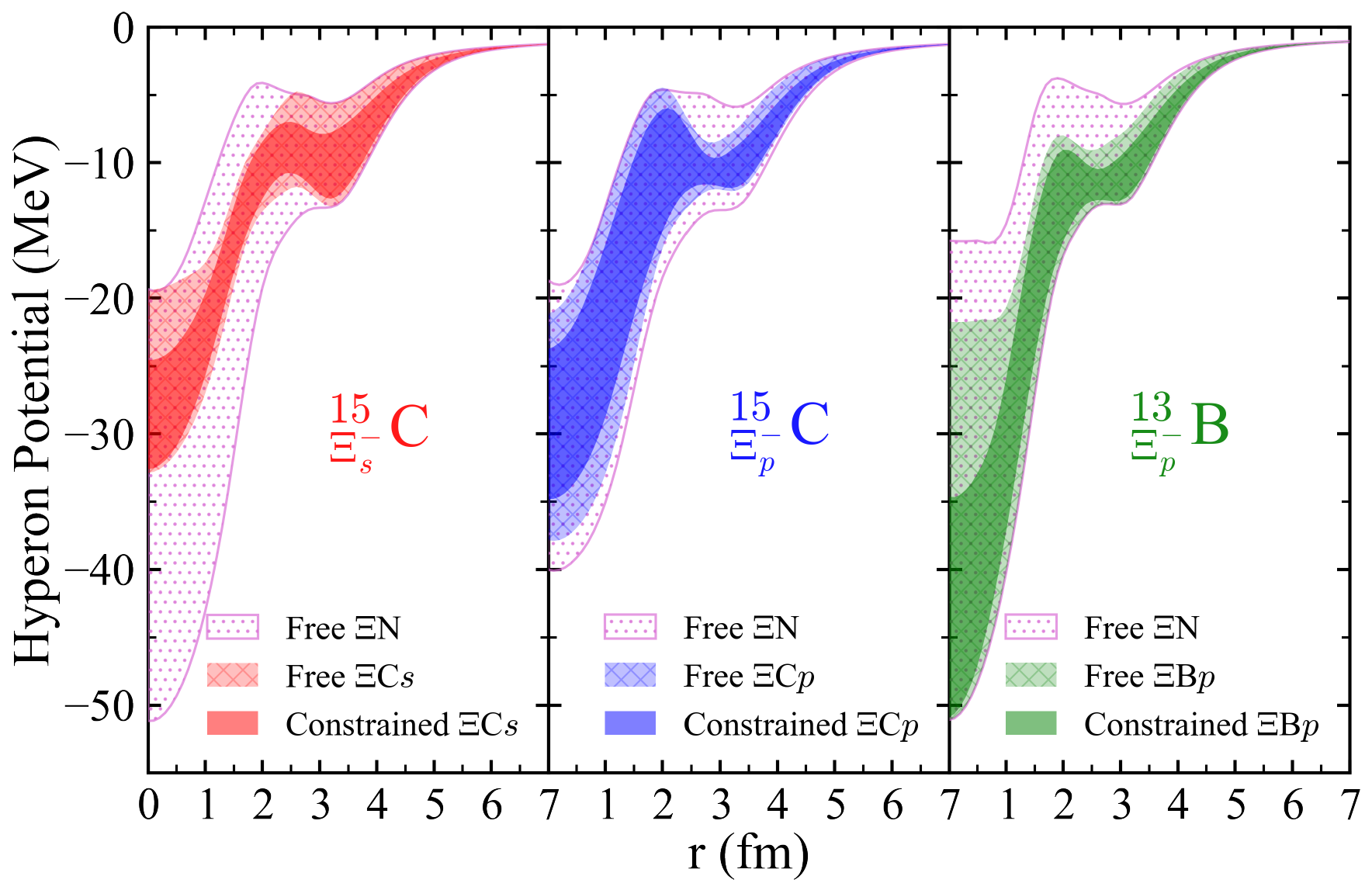}
	\caption{Using the six RMF effective Lagrangians listed in Table \ref{Tab:CouplingStrengthX}, we calculate the uncertainties of the hyperon potentials in $^{15}_{\Xi^{-}}$C and $^{13}_{\Xi^{-}}$B hypernuclei with the $\Xi^{-}$ hyperon in the $1s$ or $1p$ state, respectively. In the three panels, the light color meshed regions represent the results obtained by adopting only one kind of $\Xi N$ interaction (denoted as Free $\Xi$C$s$, Free $\Xi$C$p$, and Free $\Xi$B$p$) within the range $R_{\omega\Xi}=0.200\text{--}1.000$, while the dark color areas show the constrained distribution after considering the joint +ASTRO+NUCL$\Lambda$+NUCL$\Xi$ analysis (see Table \ref{Tab:CouplingStrength} in Sec. \ref{sec:result} for details). The pink dotted filled areas indicate the theoretical uncertainty ranges estimated by combining the three fitting strategies of $\Xi N$ interactions (denoted as Free $\Xi N$).}\label{Fig:Potential}
\end{figure}

The second strategy is based on the separation energy of $\Xi^{-}$ in the $1p$ state of $^{15}_{\Xi^{-}}$C. This deeply bound hypernucleus was first observed in the KISO event in 2015 \cite{Nakazawa2015Prog.Theor.Exp.Phys.2015.033D02}. Similar to the KINKA event, the KISO event reported two separation energy values without a definitive assignment of the $\Xi^{-}$ nuclear state. Later remeasurements in the IBUKI event, combined with theoretical analyses, ultimately attributed the smaller separation energy from KISO, $B_{\Xi^{-}}=1.03 \pm 0.18$ MeV, and the separation energy from IBUKI, $B_{\Xi^{-}}=1.27 \pm 0.21$ MeV, to the $\Xi^{-}$ in the $1p$ state. As these two values are consistent within $1\sigma$ uncertainty, Ref.~\cite{Hayakawa2021PRL126.062501} performed a weighted average under the assumption of the same initial state, yielding $B_{\Xi^{-}}=1.13 \pm 0.14$ MeV and recommending this as the experimental reference value. In this work, we adopt this central value, $B_{\Xi^{-}}=1.13$ MeV, as the target for fitting $R_{\sigma\Xi}$, and denote the resulting interaction as $\Xi$C$p$.

\begin{figure}[htbp]
	\centering
	\includegraphics[width=0.48\textwidth]{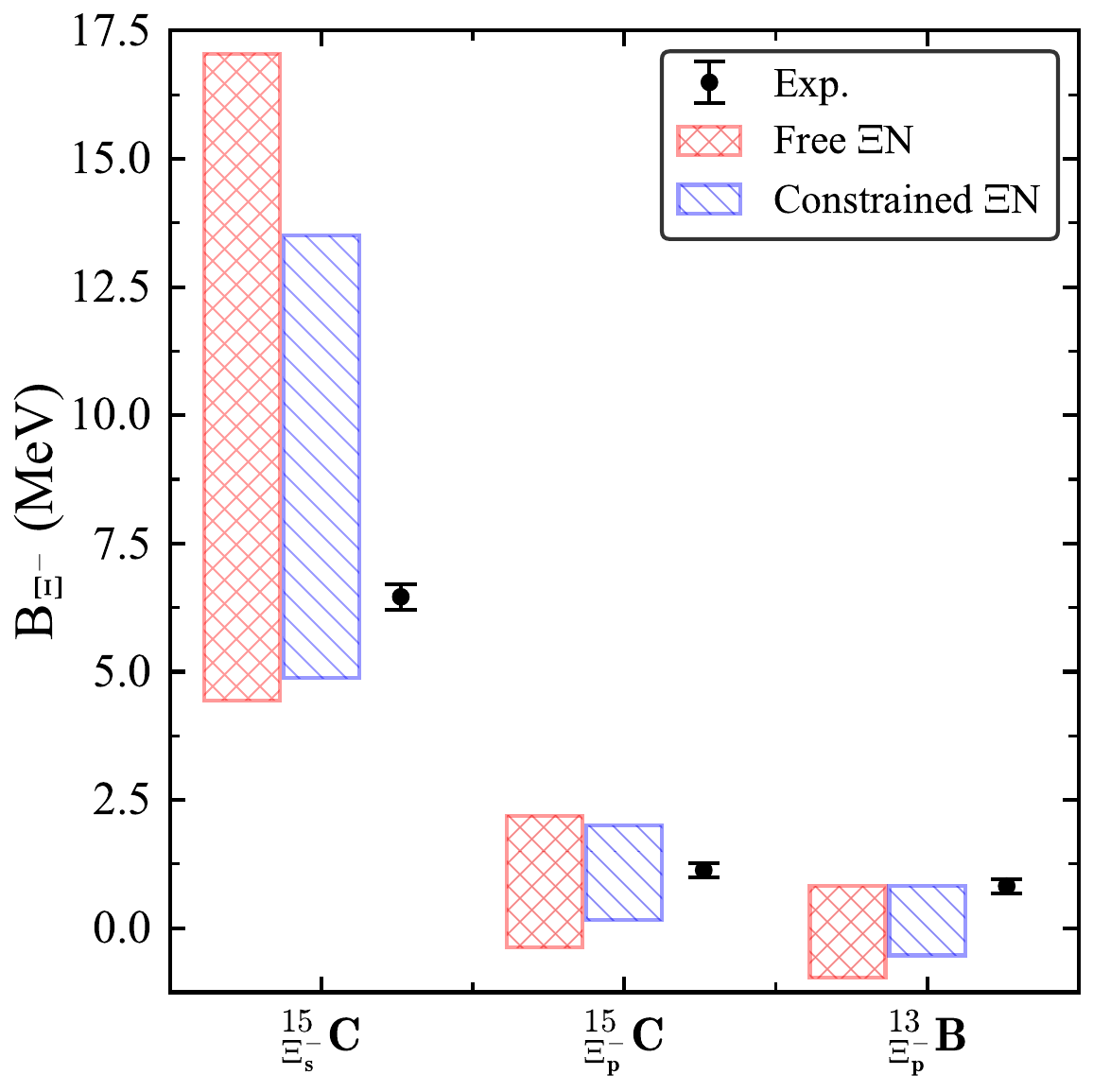}
	\caption{The hyperon separation energies $B_{\Xi^{-}}$ of $^{15}_{\Xi^{-}}$C and $^{13}_{\Xi^{-}}$B are calculated using four RMF effective Lagrangians PKDD, DD-ME2, DD-MEX, and DD-LZ1, assuming that the $\Xi^{-}$ in either the $1s$ or $1p$ state. The red bars indicate the maximum uncertainties of the separation energies resulting from the combination of the three fitting strategies of $\Xi N$ interactions (denoted as Free $\Xi$C$s$, Free $\Xi$C$p$, and Free $\Xi$B$p$) as $R_{\omega\Xi}$ varies from 0.200 to 1.000,  while blue bars denote the maximum uncertainties predicted after considering the joint +ASTRO+NUCL$\Lambda$+NUCL$\Xi$ analysis (see Table \ref{Tab:CouplingStrength} in Sec. \ref{sec:result} for details). The black error bars correspond to the experimentally measured $\Xi^{-}$ hypernuclear data.}\label{Fig:SeparationEnergy}
\end{figure}

In addition, some studies have suggested that in $^{15}_{\Xi}$C events, the $\Xi^{-}$ state in $^{14}$N may mix with the $\Xi^{0}$ state in $^{14}$C. For example, the IRRAWADDY event has been interpreted in Ref.~\cite{Friedman2023PLB837.137640} as a $^{14}$C+$\Xi^{0}_{p}$ bound state. Consequently, it has been recommended that greater attention be paid to capture events involving $^{12}$C and $^{16}$O, where the $\Xi^{-}p \leftrightarrow \Xi^{0} n$ coupling is ineffective. Notably, early emulsion experiments on the $^{13}_{\Xi^{-}}$B ($^{12}$C+$\Xi^{-}$) hypernucleus provided information on the $\Xi^{-}$ separation energy. Although no clearly bound hypernuclear state was directly observed, two independent events produced consistent results, $B_{\Xi^{-}}=0.82 \pm 0.17$ MeV and $B_{\Xi^{-}}=0.82 \pm 0.14$ MeV, both corresponding to the $\Xi^{-}_{1p}$ nuclear state. These findings have attracted considerable experimental and theoretical interest. Based on this, the third strategy adopts this value as a new fitting target for constructing the $\Xi N$ interaction. By fitting the central value $B_{\Xi^{-}}=0.82$ MeV, we determine $R_{\sigma\Xi}$ and denote the resulting interaction as $\Xi$B$p$. It should be noted that, since the experimental data do not resolve the $\Xi^{-}{1p}$ spin-orbit splitting, the fittings for both the $\Xi$C$p$ and $\Xi$B$p$ series are performed by averaging over the $\Xi^{-}$ spin doublet states with the same orbital angular momentum $l_{\Xi^{-}}$.
\begin{figure}[htbp]
	\centering
	\includegraphics[width=0.48\textwidth]{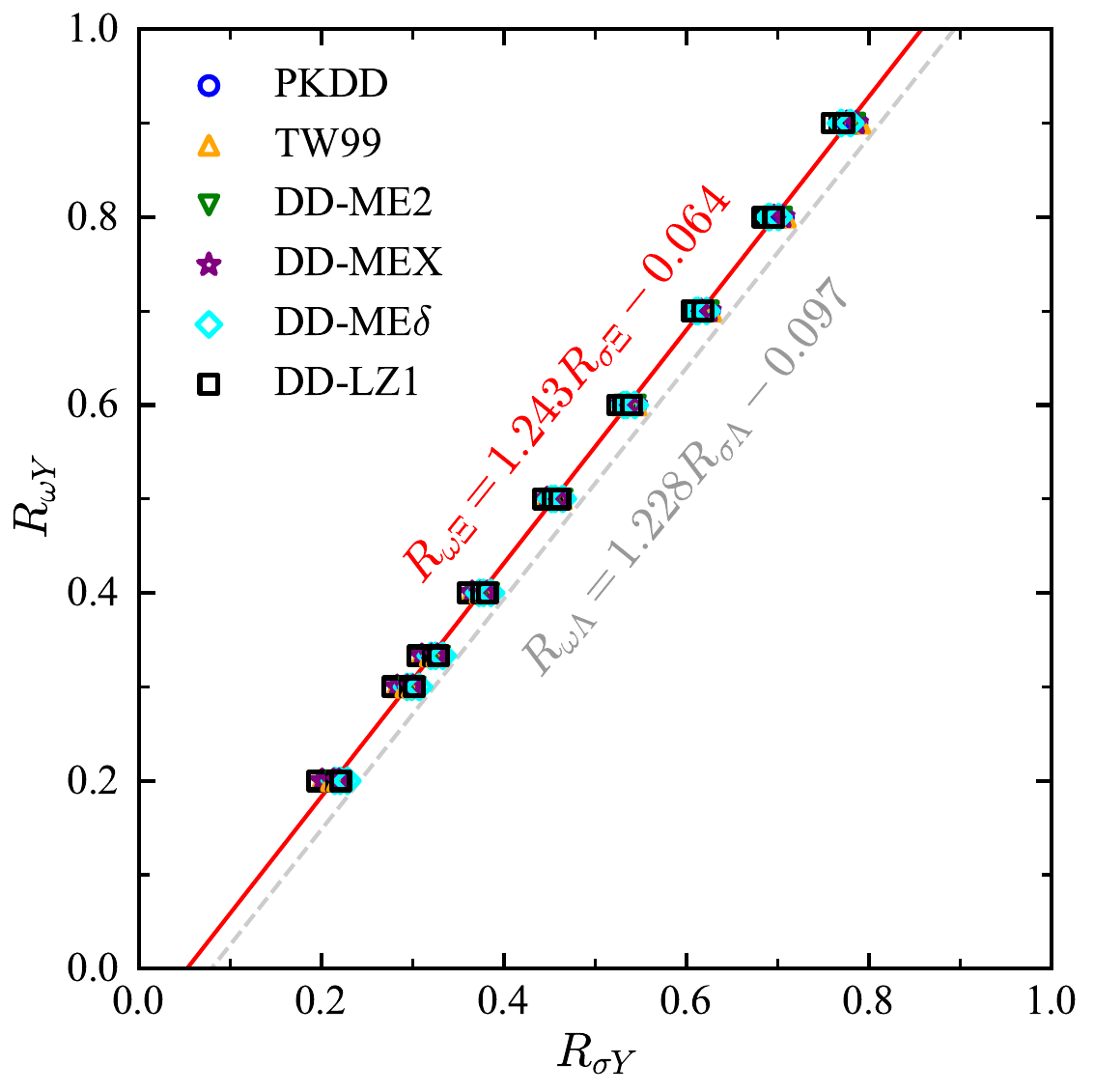}
	\caption{The correlation between $R_{\sigma\Xi}$ and $R_{\omega\Xi}$ for the effective interactions listed in Table \ref{Tab:CouplingStrengthX} is presented. A linear fit to the data yields the relation $R_{\omega\Xi}=1.243R_{\sigma\Xi}-0.064$, which is illustrated by the red line. Results from $\Lambda$ hypernuclei (from Ref. \cite{Rong2021PRC104.054321}) are also shown as a gray dashed line.}\label{Fig:Linear}
\end{figure}

The $\Xi N$ effective interactions determined by the three fitting strategies discussed above are summarized in Table~\ref{Tab:CouplingStrengthX}. To more clearly illustrate their impact on hypernuclear structure, we calculated the hyperon potentials of $^{15}_{\Xi^{-}}$C and $^{13}_{\Xi^{-}}$B using the six RMF effective Lagrangians in Table~\ref{Tab:CouplingStrengthX}, assuming the $\Xi^{-}$ occupies either the $1s$ or $1p$ state, as shown in Fig.~\ref{Fig:Potential}. In the three panels, the light red, light blue, and light green meshed regions correspond to results derived from the $\Xi$C$s$, $\Xi$C$p$, and $\Xi$B$p$ interactions, respectively, within the range $R_{\omega\Xi}=0.200\text{--}1.000$. The pink dotted areas indicates the estimated theoretical uncertainty band obtained by combining the three interactions. Table \ref{Tab:CouplingStrengthX} does not list the results for the three $\Xi N$ interactions at $R_{\omega\Xi}=1.000$. However, since the Bayesian constraint analysis suggests that $R_{\omega\Xi}$ can exceed $0.900$ for certain Lagrangians, we additionally performed calculations in the range $R_{\omega\Xi}=0.900\text{--}1.000$ to more comprehensively capture the theoretical uncertainties. Since current theoretical models are not yet capable of consistently reproducing all available experimental data, the $\Xi N$ interaction is typically constrained only by specific hypernuclear states in order to reduce theoretical uncertainties. Accordingly, the $\Xi$C$s$ interaction, fitted to the $1s$ state of $^{15}_{\Xi^{-}}$C, yields smaller uncertainties in the high-density nuclear core region, while the $\Xi$C$p$ and $\Xi$B$p$ interactions, fitted to the $1p$ states of $^{15}_{\Xi^{-}}$C and $^{13}_{\Xi^{-}}$B respectively, yield smaller uncertainties in the low-density nuclear surface regions. Overall, the combined use of these three interactions provides a reliable estimate of the model uncertainties under the constraints of existing data.

We further calculated the hyperon separation energies of three representative hypernuclear states $^{15}_{\Xi^{-}_{s}}$C, $^{15}_{\Xi^{-}_{p}}$C, and $^{13}_{\Xi^{-}_{p}}$B. For consistency with the Bayesian analysis, the calculations employed only four RMF effective Lagrangians PKDD, DD-ME2, DD-MEX, and DD-LZ1, with the corresponding results shown in Fig. \ref{Fig:SeparationEnergy}. The red bars denote the maximum uncertainty ranges of the predicted separation energies, obtained by combining the three sets of $\Xi N$ effective interactions ($\Xi$C$s$, $\Xi$C$p$, and $\Xi$B$p$) with the four Lagrangians, while varying $R_{\omega\Xi}$ from $0.200$ to $1.000$. The black error bars indicate the experimental data. We also verified the results using TW99 and DD-ME$\delta$, confirming that the current calculations already encompass their predicted ranges. As shown, when the $\Xi N$ interaction has large uncertainties, the resulting hyperon separation energies exhibit pronounced model dependence and significant deviations from experimental values. This underscores the urgent need for more effective approaches to constrain the $\Xi N$ interaction.

Following the idea in Ref.~\cite{Sun2023APJ942.55}, we first explored the possible correlation between $R_{\sigma\Xi}$ and $R_{\omega\Xi}$. Using the data in Table~\ref{Tab:CouplingStrengthX}, we plotted $R_{\omega\Xi}$ as a function of $R_{\sigma\Xi}$, shown in Fig.~\ref{Fig:Linear}, which clearly demonstrates a pronounced linear correlation between the two parameters. A linear fit to all data points yields the following relation:
\begin{align}\label{eq:LinearRelationship}
	R_{\omega\Xi} = 1.243R_{\sigma\Xi} - 0.064
\end{align}
The result is shown as the red line in Fig. \ref{Fig:Linear}. For reference, the scalar-vector coupling relation derived from $\Lambda$ hypernuclei, taken from Ref. \cite{Rong2021PRC104.054321}, is also shown as a gray dashed line. Since the linear empirical relation for $\Xi$ hypernuclei will be incorporated into the subsequent Bayesian analysis, and the likelihood function construction relies on the statistical uncertainty of $R_{\sigma\Xi}$, denoted as $\sigma_{R_{\sigma\Xi}}$, it is essential to assess the uncertainty in $R_{\sigma\Xi}$ for $\Xi N$ effective interactions.

In the present work, the $\Xi N$ effective interaction was constrained solely by the hyperon separation energy of a specific $\Xi$ hypernucleus. Consequently, unlike in Ref. \cite{Rong2021PRC104.054321}, where parameter uncertainties were directly estimated through statistical analyses, such an approach is not applicable here, as meaningful error estimation requires at least two data points \cite{Brandt2014statistical, Dobaczewski2014JPG41.074001}. As an alternative, we take the linear relation shown in Fig. \ref{Fig:Linear} as the central value and define the statistical error as the deviation of $R_{\sigma\Xi}$, predicted by each RMF effective Lagrangian, from this central value at the 95\% prediction interval bounds. Within this framework, $\sigma_{R_{\sigma\Xi}}=0.05$ provides a reasonable estimate of the uncertainty. However, since only the central value was used to fit the $\Xi N$ interaction, and experimental uncertainties, which contribute an additional $\Delta\sigma_{R_{\sigma\Xi}} \approx 0.01$, as well as potential systematic biases from other RMF models are not fully captured, we conservatively extend the error to $\sigma_{R_{\sigma\Xi}} = 0.08$. This choice more adequately reflects the statistical uncertainty across different models. Indeed, as illustrated in Figure 9 of Ref. \cite{Sun2023APJ942.55}, modest variations in $\sigma_{R_{\sigma\Xi}}$ do not affect the conclusions of the Bayesian analysis. Therefore, in subsequent inference, we apply tighter constraints on the $YN$ interactions using the linear relation obtained in this section together with the assessed statistical error of $R_{\sigma\Xi}$.

\section{Bayesian inference}  \label{sec:analysis} 

Given a model with parameters $\bm{\theta}$ and observed data $\bm{D}$, the posterior probability is obtained via Bayes' theorem
\begin{equation}
	P(\bm{\theta}|\bm{D}) = \frac{P(\bm{D}|\bm{\theta})P(\bm{\theta})}{\int P(\bm{D}|\bm{\theta})P(\bm{\theta})\rm{d}\bm{\theta}},
\end{equation}
where $P(\bm{\theta})$ denotes the prior probability of the parameter set $\bm{\theta}$. The likelihood function $P(\bm{D}|\bm{\theta})$ is given by the product of individual likelihoods $P_i({\bm{d}}_{i}|\bm{\theta})$ for each observation ${\bm{d}}_{i}\in\bm{D}$. Below, we discuss in detail the prior and likelihood used in our analysis

\subsection{Dataset and likelihood}

In this work, we employ four types of experimental data: mass-radius measurements of X-ray pulsars (PSR J0030+0451 and PSR J0740+6620) from NICER (labeled NICER), the tidal deformability measurement (GW170817) from LIGO/Virgo (labeled GW170817), and laboratory measurements of single-$\Lambda$ (labeled NUCL$\Lambda$) and single-$\Xi$ hypernuclei (labeled NUCL$\Xi$). See Ref. \cite{Sun2023APJ942.55} for details on the likelihood functions $P_{\mathrm{NUCL}\Lambda}(\bm{d}_{\mathrm{NUCL}\Lambda}|\bm{\theta})$, $P_{\mathrm{NICER}}(\bm{d}_{\mathrm{NICER}}|\bm{\theta})$ and $P_{\rm GW}(\bm{d}_{\rm{GW}}|\bm{\theta})$. The only difference is that we replace the previously adopted PSR J0030+0451 data from Ref. \cite{Riley2019APJL887.L21} to Ref. \cite{Vinciguerra2024APJ961.62}, and update the PSR J0740+6620 data from Ref. \cite{Riley2021APJL918.L27} to those in Ref. \cite{Salmi2024APJ974.294}. 
Then, only the likelihood function for the $\Xi$ hypernuclear data is presented in detail below.

There is an excellent linear correlation between the ratio of coupling strengths $R_{\sigma\Xi}$ and $R_{\omega\Xi}$ for $\Xi$ hypernuclei, similar to $\Lambda$ hypernuclei, as shown in Fig. \ref{Fig:Linear}. On top of our previous study solely based on $\Lambda$ hypernuclei \cite{Sun2023APJ942.55}, we incorporate this correlation as a new nuclear physics constraint in our Bayesian analysis, with the likelihood function constructed as follows
\begin{align}
	P_{\mathrm{NUCL}\Xi}(\bm{d}_{\mathrm{NUCL}\Xi}|\bm{\theta})=\exp\left[-\frac{1}{2}\frac{(R_{\sigma\Xi}-\bar{ R}_{\sigma\Xi})^2}{\sigma^{2}_{R_{\sigma\Xi}}}\right],
\end{align}
where $\bar R_{\sigma\Xi}=(R_{\omega\Xi}+0.064)/1.243$ is derived from the empirical relationship from Eq. (\ref{eq:LinearRelationship}), and the corresponding standard deviation is given by $\sigma_{R_{\sigma\Xi}}=0.08$ as discussed above.

\subsection{Model parameters and priors} \label{Sec:test}

In the present work, the model parameters are categorized into three groups:

1) The EOS parameters $\bm{\theta}_{\rm EOS} =\{R_{\sigma Y},R_{\omega Y}\}$ ($Y=\Lambda, \Xi$). To ensure consistency, we follow Ref. \cite{Sun2023APJ942.55}, assuming that the $\Lambda$ hyperon-meson couplings are weaker than those of nucleons and the ratios of the coupling strengths, $R_{\sigma Y}$ and $R_{\omega Y}$, follow uniform distributions, i.e., $R_{\sigma Y} \sim U[0, 1]$ and $R_{\omega Y} \sim U[0, 1]$. It should be noted that $\Sigma$ hyperon, an essential component of the baryon octet, has so far only been observed in the form of $^{4}_{\Sigma}$He \cite{Hayano1989PLB231.355, Nagae1998PRL80.1605}. The $\Sigma$ hyperon potential is generally considered to be repulsive, making it difficult to extract $\Sigma N$ interaction information from hypernuclear systems in the same way as with $\Lambda$ or $\Xi$ hypernuclei \cite{Hayano1989PLB231.355, Mares1995NPA594.311, Nagae1998PRL80.1605}. In the present work, we adopt the method employed in previous studies and determine the $\Sigma N$ interaction in symmetric nuclear matter by fitting the $\Sigma$ potential at saturation density to a value of $U_{\Sigma}=+30$ MeV \cite{Friedman2007Phys.Rep.452.89, Ishizuka2008JPG35.085201, Fortin2017PRC95.065803}.

2) When considering the NICER measurements, we include the central energy density of pulsar $j$, $\varepsilon_{c,j}$, into the parameter set to obtain its mass and radius, $M=M(\bm{\theta}_{\rm{EOS}};\varepsilon_{c,  j})$ and $R=R(\bm{\theta}_{\rm{EOS}};\varepsilon_{c, j})$. 
We assign physically reasonable and wide enough uniform priors for the central energy density as $\varepsilon_{c}\sim U[0.3\times10^{15},1\times10^{15}],\rm{g/cm^3}$ for PSR J0030+0451, and as $\varepsilon_{c}\sim U[0.6\times10^{15},3\times10^{15}], \rm{g/cm^3}$ for PSR J0740+6620.

3) The gravational wave parameters are the chirp mass $\mathcal{M}$ and the mass ratio $q$, while the tidal deformabilities of two components are determined by the EOS and their respective masses, i.e., $\Lambda_{1}(\bm{\theta}_{\rm{EOS}}; M_1)$ and $\Lambda_{2}(\bm{\theta}_{\rm{EOS}}; M_2)$. We adopt uniform priors for the chirp mass $\mathcal{M}\sim U[1.18, 1.21]~M_{\odot}$ and the mass ratio $q\sim U[0.5, 1]$.

With the priors and likelihoods specified, we sample from the posterior distribution using the Python-based \textsf{bilby}~\cite{Ashton2019ascl.soft01011A} and \textsf{pymultinest}~\cite{Buchner2016ascl.soft06005B} packages. We conduct four main tests to examine the impact of individual astrophysical and laboratory data on the $\Lambda$ and $\Xi$ coupling strengths, namely:\\
(i)+ASTRO: where we consider the constraints from NICER for PSR J0030+0451 and PSR J0740+6620, along with the GW170817 data;\\
(ii)+ASTRO+NUCL$\Lambda$: where we consider both the constraints in (i) and the $\Lambda$ hypernuclei ones;\\
(iii)+ASTRO+NUCL$\Xi$: where we consider both the constraints in (i) and the $\Xi$ hypernuclei ones;\\
(iv)+ASTRO+NUCL$\Lambda$+NUCL$\Xi$: where we consider the constraints from (i) along with those from $\Lambda$ and $\Xi$ hypernuclei.

\section{Results and discussion} \label{sec:result}

\begin{figure}[htbp]
	\centering
	\includegraphics[width=0.48\textwidth]{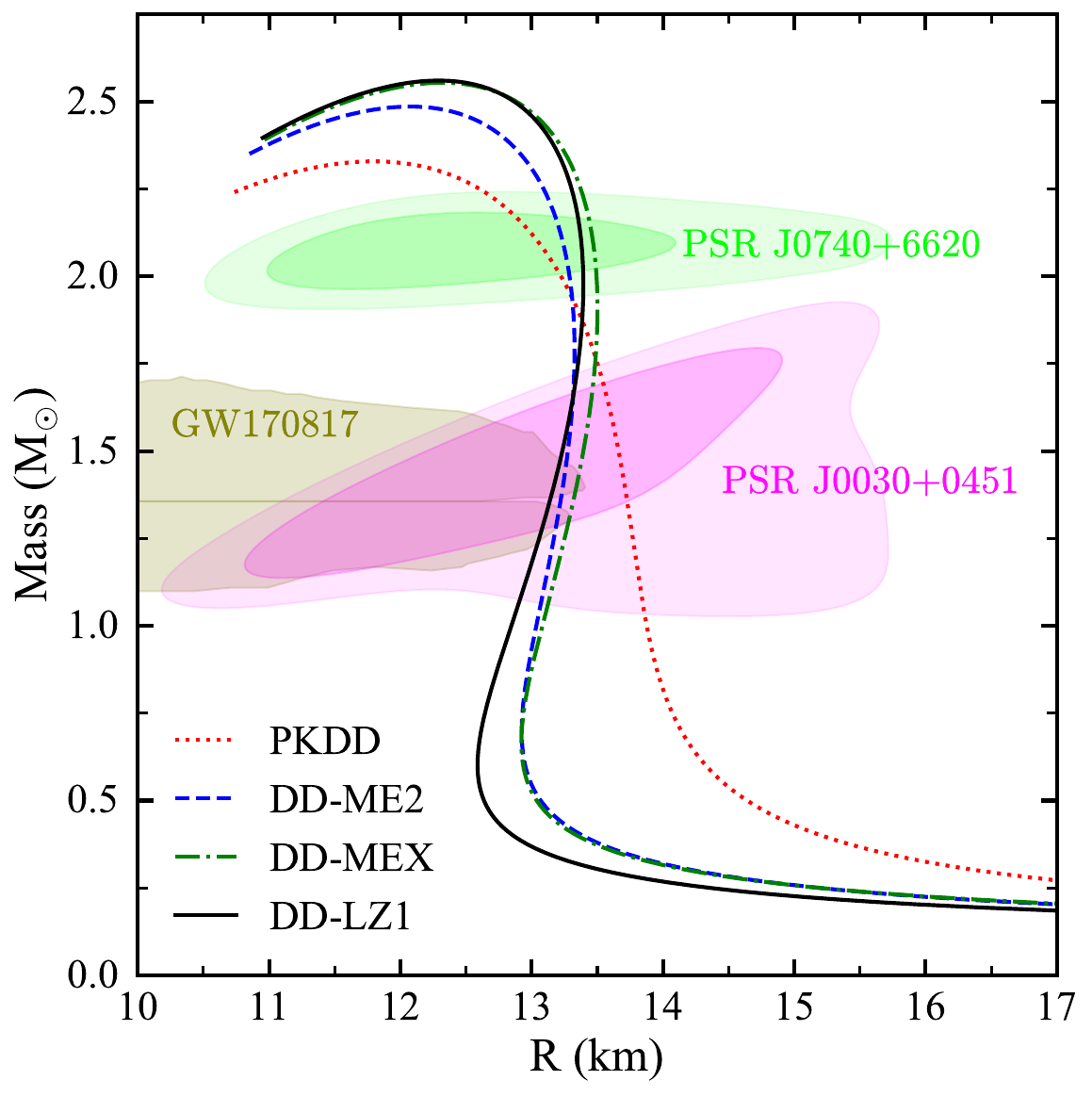}
	\caption{The mass–radius relations of NSs are calculated using various RMF effective interactions. For comparison, we include the mass–radius measurements from the NICER for PSR J0030+0451 (from PDT-U)~\cite{Vinciguerra2024APJ961.62} and PSR J0740+6620~\cite{Salmi2024APJ974.294}, both shown at the 68\% and 95\% confidence levels, as well as the binary tidal deformability constraints from GW170817 reported by LIGO/Virgo~\cite{Abbott2017PRL119.161101}.}\label{Fig:MR}
\end{figure}

Before discussing the properties of hyperon stars, we first examine the behavior of several sets of effective Lagrangians employed for constructing the $\Xi N$ effective interaction in NS matter. Note that the TW99 and DD-ME$\delta$ effective Lagrangians produce soft EOSs, and including hyperon degrees of freedom further softens them. Consequently, their predicted maximum hyperon star masses fall well below the observational lower limit of approximately $2M_{\odot}$, leading us to exclude them from further analysis. We thus focus on the remaining relativistic density functionals, and show the corresponding NS mass-radius relations in Fig. \ref{Fig:MR}. Following Ref. \cite{Sun2023APJ942.55}, the crust structure is described using the quantum calculations of Ref. \cite{Negele1973NPA207.298} for the intermediate-density regime ($0.001~\rm{fm}^{-3}<\rho<0.08~\rm{fm}^{-3}$), and the theoretical framework developed by Ref. \cite{Baym1971APJ170.299} for the outer crust ($\rho<0.001~\rm{fm}^{-3}$). For comparison, the figure also includes mass-radius constraints inferred from the GW170817 tidal deformability measurements by LIGO/Virgo \cite{Abbott2017PRL119.161101}, as well as NICER observations for PSR J0030+0451 \cite{Vinciguerra2024APJ961.62} and PSR J0740+6620 \cite{Salmi2024APJ974.294}. As shown in Fig. \ref{Fig:MR}, the RMF effective Lagrangians in consideration are broadly consistent with current observational constraints, with PKDD predicting relatively large NS radii. Among them, DD-LZ1 yields the highest NS mass while remaining consistent with observational constraints, and is therefore selected as the representative model for subsequent analysis.

\subsection{Hyperon-nucleon interactions}

\begin{figure}[htbp]
	\centering
	\includegraphics[width=0.48\textwidth]{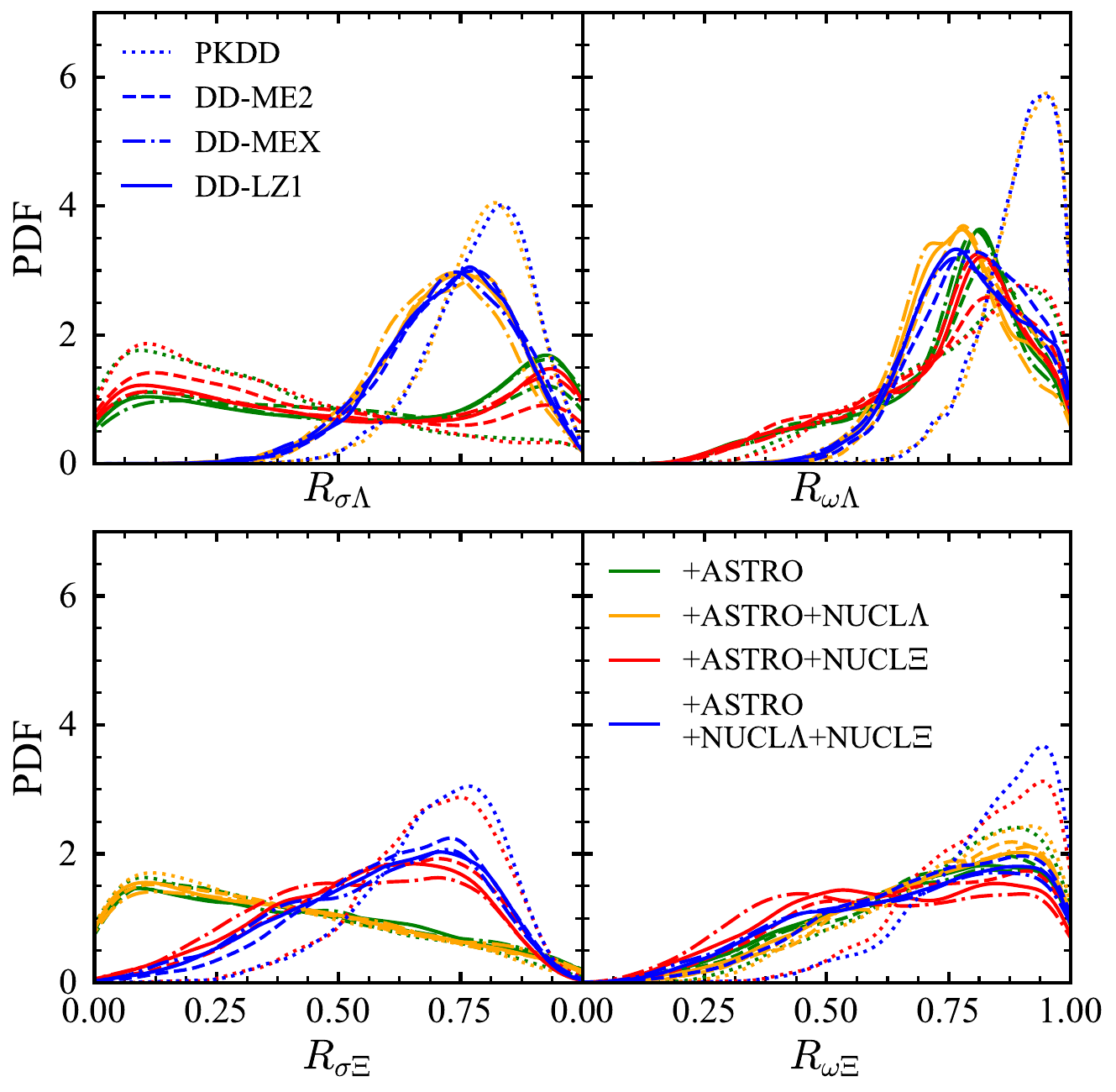}
	\caption{Posterior PDFs of $R_{\sigma\Lambda}$ (top-left panel), $R_{\omega\Lambda}$ (top-right panel), $R_{\sigma\Xi}$ (bottom-left panel), and $R_{\omega\Xi}$ (bottom-right panel) for the scalar and vector couplings between $\Lambda N$ and $NN$, or $\Xi N$ and $NN$ interactions, based on the RMF models adopted in this work. This analysis incorporates constraints from GW170817 and NICER (PSR J0030+0451 and PSR J0740+6620). Results are shown under different conditions: considering only astrophysical observational constraints (green curves), adding the empirical $R_{\sigma\Lambda}$-$R_{\omega\Lambda}$ relation constrained by single-$\Lambda$ hypernuclear data (orange curves), adding the empirical $R_{\sigma\Xi}$-$R_{\omega\Xi}$ relation constrained by single-$\Xi$ hypernuclear data (red curves), and incorporating both empirical relations from single-$\Lambda$ and single-$\Xi$ hypernuclear data (blue curves).}\label{Fig:PDF}
\end{figure}

\begin{figure}[htbp]
	\centering
	\includegraphics[width=0.48\textwidth]{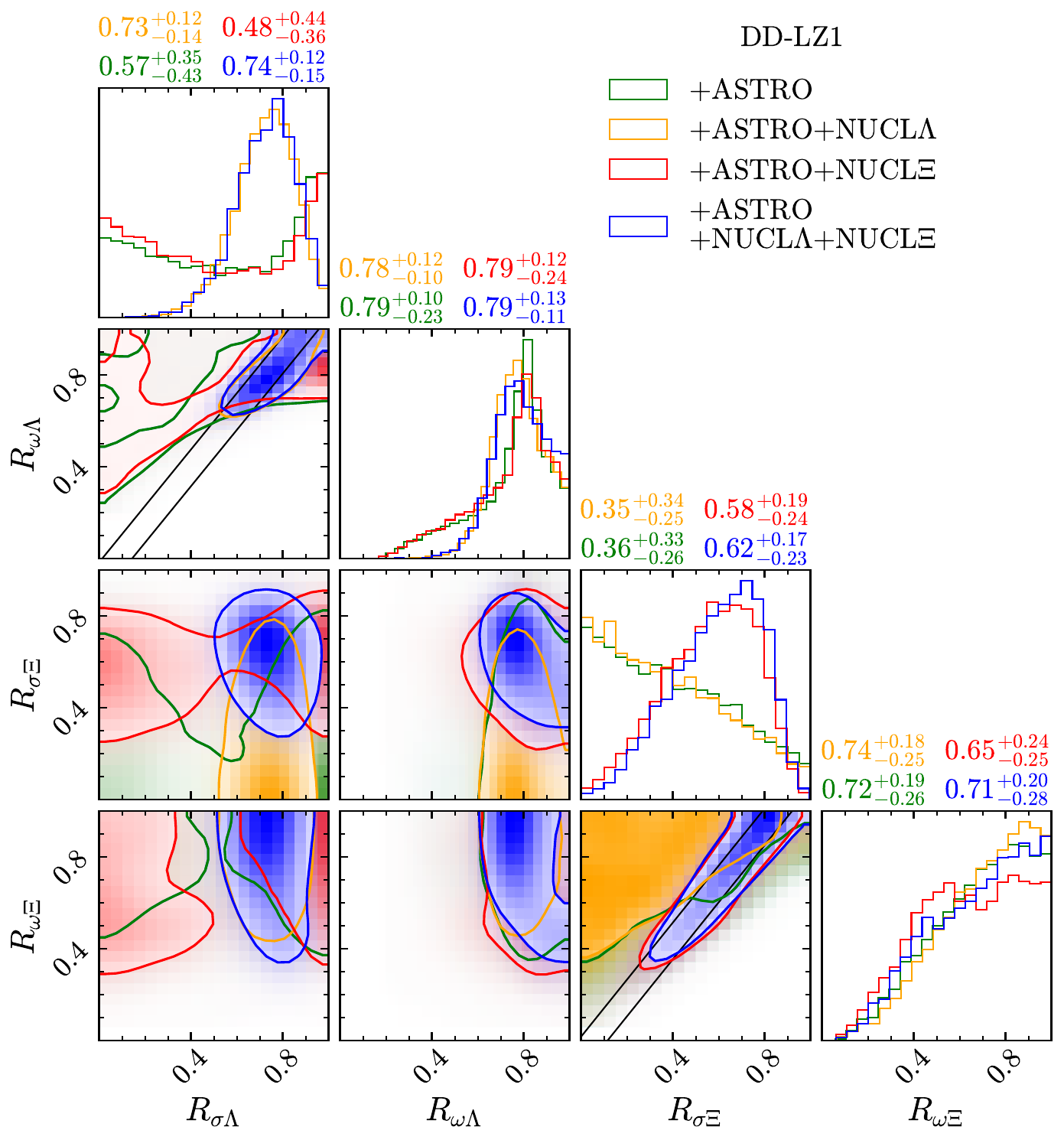}
	\caption{Shown are the posterior distributions of $R_{\sigma\Lambda}$, $R_{\omega\Lambda}$, $R_{\sigma\Xi}$, and $R_{\omega\Xi}$. Green indicates results constrained only by astronomical observations; orange includes additional empirical constraints from $\Lambda$ hypernuclei; red includes constraints from $\Xi$ hypernuclei; and blue includes both $\Lambda$ and $\Xi$ hypernuclear constraints simultaneously. The empirical laboratory $R_{\sigma\Lambda}$-$R_{\omega\Lambda}$ relation, derived from $\Lambda$ separation energy measurements in single-$\Lambda$ hypernuclei, and the $R_{\sigma\Xi}$-$R_{\omega\Xi}$ relation, inferred from $\Xi$ separation energy measurements in single-$\Xi$ hypernuclei, are indicated by the two black lines. Contours show the 68\% credible regions obtained with the effective interaction DD-LZ1.}\label{Fig:CornerAll}
\end{figure}

Table \ref{Tab:CouplingStrength} summarizes the most probable values of the hyperon coupling strength ratios $R_{\sigma Y}$ and $R_{\omega Y}$, along with their 68\% confidence boundaries, inferred from four different tests that combine astrophysical and nuclear physics constraints. Based on this, we adopt the $R_{\omega \Xi}$ constraint obtained from the combined +ASTRO+NUCL$\Lambda$+NUCL$\Xi$ analysis to illustrate the hyperon potentials for three representative hypernuclear states under the $\Xi$C$s$, $\Xi$C$p$ and $\Xi$B$p$ interactions, as shown by the red, blue, and green filled regions in Fig. \ref{Fig:Potential}. Compared with the unconstrained $\Xi N$ interactions, the distributions of the hyperon potentials are significantly narrowed. Furthermore, using the four effective Lagrangians listed in the table, we evaluate the maximum uncertainties in the separation energies of the three representative hypernuclear states predicted by the three $\Xi N$ interaction sets. The results, shown as blue bars in Fig. \ref{Fig:SeparationEnergy}, demonstrate a substantial reduction in uncertainty compared with the unconstrained $\Xi N$ interactions. These findings clearly highlight the effectiveness of Bayesian analysis in constraining hyperon interactions. The corresponding posterior probability density functions (PDFs) for each RMF effective Lagrangian are shown in Fig. \ref{Fig:PDF}. For the DD-LZ1 model, additional results under various test conditions are presented in Fig. \ref{Fig:CornerAll}, highlighting the individual effects of astrophysical versus nuclear constraints.

\begin{table*}[hbpt]
	\footnotesize 
	\centering
	\caption{Most probable values of $R_{\sigma\Lambda}$, $R_{\omega\Lambda}$, $R_{\sigma\Xi}$, and $R_{\omega\Xi}$, with their 68\% credible intervals.}\label{Tab:CouplingStrength}
	\renewcommand{\arraystretch}{2.0}
	\doublerulesep 0.1pt \tabcolsep 4.5pt
	\begin{tabular}{lccccccccclcc}
		\hline\hline
		&                      &                           &                           &  & \multicolumn{2}{c}{+ASTRO}                            &  & \multicolumn{2}{c}{+ASTRO}                            &  & \multicolumn{2}{c}{+ASTRO}                            \\ [-2.5mm]
		\multirow{2}{*}{}       & \multirow{2}{*}{$Y$} & \multicolumn{2}{c}{+ASTRO}                            &  & \multicolumn{2}{c}{+NUCL$\Lambda$}                    &  & \multicolumn{2}{c}{+NUCL$\Xi$}                        &  & \multicolumn{2}{c}{+NUCL$\Lambda$+NUCL$\Xi$}          \\ \cline{3-4} \cline{6-7} \cline{9-10} \cline{12-13} 
		&                      & $R_{\sigma Y}$            & $R_{\omega Y}$            &  & $R_{\sigma Y}$            & $R_{\omega Y}$            &  & $R_{\sigma Y}$            & $R_{\omega Y}$            &  & $R_{\sigma Y}$            & $R_{\omega Y}$            \\ \hline
		\multirow{2}{*}{PKDD}   & $\Lambda$            & 0.289$^{+0.342}_{-0.213}$ & 0.812$^{+0.133}_{-0.214}$ &  & 0.801$^{+0.090}_{-0.108}$ & 0.911$^{+0.060}_{-0.085}$ &  & 0.288$^{+0.339}_{-0.204}$ & 0.816$^{+0.132}_{-0.210}$ &  & 0.806$^{+0.084}_{-0.119}$ & 0.911$^{+0.059}_{-0.088}$ \\
		& $\Xi$                & 0.315$^{+0.330}_{-0.224}$ & 0.784$^{+0.145}_{-0.230}$ &  & 0.305$^{+0.335}_{-0.212}$ & 0.790$^{+0.145}_{-0.229}$ &  & 0.700$^{+0.122}_{-0.145}$ & 0.836$^{+0.116}_{-0.164}$ &  & 0.712$^{+0.114}_{-0.156}$ & 0.859$^{+0.099}_{-0.164}$ \\ \hline
		\multirow{2}{*}{DD-ME2} & $\Lambda$            & 0.486$^{+0.419}_{-0.358}$ & 0.796$^{+0.110}_{-0.235}$ &  & 0.736$^{+0.121}_{-0.139}$ & 0.792$^{+0.123}_{-0.103}$ &  & 0.371$^{+0.483}_{-0.268}$ & 0.788$^{+0.135}_{-0.245}$ &  & 0.744$^{+0.120}_{-0.143}$ & 0.808$^{+0.115}_{-0.115}$ \\
		& $\Xi$                & 0.336$^{+0.348}_{-0.230}$ & 0.736$^{+0.181}_{-0.261}$ &  & 0.331$^{+0.341}_{-0.232}$ & 0.757$^{+0.167}_{-0.228}$ &  & 0.603$^{+0.179}_{-0.224}$ & 0.698$^{+0.213}_{-0.255}$ &  & 0.633$^{+0.158}_{-0.205}$ & 0.743$^{+0.178}_{-0.239}$ \\ \hline
		\multirow{2}{*}{DD-MEX} & $\Lambda$            & 0.590$^{+0.346}_{-0.426}$ & 0.787$^{+0.102}_{-0.204}$ &  & 0.714$^{+0.125}_{-0.136}$ & 0.764$^{+0.115}_{-0.105}$ &  & 0.499$^{+0.421}_{-0.373}$ & 0.790$^{+0.115}_{-0.212}$ &  & 0.730$^{+0.125}_{-0.146}$ & 0.786$^{+0.126}_{-0.116}$ \\
		& $\Xi$                & 0.358$^{+0.350}_{-0.263}$ & 0.716$^{+0.203}_{-0.258}$ &  & 0.360$^{+0.335}_{-0.256}$ & 0.752$^{+0.174}_{-0.247}$ &  & 0.550$^{+0.216}_{-0.233}$ & 0.618$^{+0.269}_{-0.258}$ &  & 0.612$^{+0.171}_{-0.244}$ & 0.706$^{+0.202}_{-0.284}$ \\ \hline
		\multirow{2}{*}{DD-LZ1} & $\Lambda$            & 0.573$^{+0.350}_{-0.428}$ & 0.791$^{+0.105}_{-0.227}$ &  & 0.731$^{+0.124}_{-0.142}$ & 0.775$^{+0.125}_{-0.105}$ &  & 0.478$^{+0.441}_{-0.358}$ & 0.787$^{+0.116}_{-0.239}$ &  & 0.739$^{+0.119}_{-0.146}$ & 0.787$^{+0.132}_{-0.112}$ \\
		& $\Xi$                & 0.364$^{+0.334}_{-0.262}$ & 0.717$^{+0.191}_{-0.262}$ &  & 0.345$^{+0.337}_{-0.245}$ & 0.740$^{+0.179}_{-0.248}$ &  & 0.580$^{+0.188}_{-0.235}$ & 0.650$^{+0.241}_{-0.249}$ &  & 0.617$^{+0.174}_{-0.233}$ & 0.713$^{+0.201}_{-0.278}$ \\
		\hline\hline
	\end{tabular}
\end{table*}

As shown in Fig. \ref{Fig:PDF}, both $\Lambda$ and $\Xi$ hyperons exhibit similar trends in their coupling strength ratios when nuclear physics constraints are absent (see green curves). Specifically, $R_{\sigma Y}$ tends toward smaller values, while $R_{\omega Y}$ favors larger ones. This behavior arises because the inclusion of hyperons significantly softens the EOS, reducing the predicted maximum mass of hyperon stars. However, astronomical observations impose strict limits on the maximum mass of NSs, thereby requiring a reduced hyperon content in their interiors. To satisfy this constraint, the repulsive component of the $YN$ interaction must be enhanced or the attractive component weakened, achieved by increasing $R_{\omega Y}$ or decreasing $R_{\sigma Y}$. Consequently, for hyperons such as $\Lambda$ and $\Xi$, which experience attractive potentials, the posterior probability distributions follow similar trends. Once nuclear physics constraints are imposed (either on $\Lambda$ or $\Xi$ hyperons), the strong linear correlation between $R_{\sigma Y}$ and $R_{\omega Y}$ forces $R_{\sigma Y}$ to shift toward larger values, thereby enhancing the attractive interaction. To satisfy the constraints imposed by astronomical observations, $R_{\omega Y}$ increases accordingly. This leads to a more concentrated posterior distribution and a notable reduction in the parameter space, as shown in Table \ref{Tab:CouplingStrength}.

A more detailed analysis reveals that when nuclear physics constraints are imposed on only one type of hyperon, the coupling strength ratio distribution of the other type remains largely unchanged, highlighting the dominant role of astronomical observations in this case. This conclusion is clearly evidenced by comparing the green curves with the red (or orange) ones in Figs. \ref{Fig:PDF} and \ref{Fig:CornerAll}. Notably, when nuclear physics constraints are simultaneously applied to both types of hyperons within a Bayesian inference framework, the allowed ranges for the scalar-to-vector coupling strength ratios shrink significantly for both types. However, the parameter space associated with the $\Xi$ hyperon still exhibits greater uncertainty. This is evident either from the parameter ranges in the last column of Table \ref{Tab:CouplingStrength} or by comparing the probability density distributions of the coupling strength ratios for $\Lambda$ and $\Xi$ hyperons in Figs. \ref{Fig:PDF} and \ref{Fig:CornerAll}. The underlying reason for this difference lies in the fact that $\Lambda$ hyperons possess both a stronger attractive potential and a lighter rest mass than $\Xi$ hyperons, making them more influential in determining the NS EOS. As a result, their interaction parameters are subject to tighter constraints from astronomical observations.

Furthermore, combining the empirical relationship between $R_{\sigma Y}$ and $R_{\omega Y}$ derived from laboratory data with astrophysical constraints leads to a slight rotation of their linear correlation, as indicated by the two black lines in Fig. \ref{Fig:CornerAll}, shifting toward lower values of $R_{\omega Y}$. At the same time, the upper-left corner of Fig. \ref{Fig:CornerAll} clearly shows that hypernuclear experimental constraints effectively exclude the small-$R_{\sigma Y}$ region. Finally, it should be emphasized that when the empirically derived linear $\Xi N$ effective interaction is employed to calculate the hyperon separation energies of several representative hypernuclei, the theoretical predictions exhibit significantly better agreement with experimental data when both $R_{\sigma Y}$ and $R_{\omega Y}$ assume relatively large values. This finding is fully consistent with the conclusions drawn in the preceding analysis.

\subsection{EOS and Hyperon Star Properties}

\begin{figure}[htbp]
	\centering
	\includegraphics[width=0.48\textwidth]{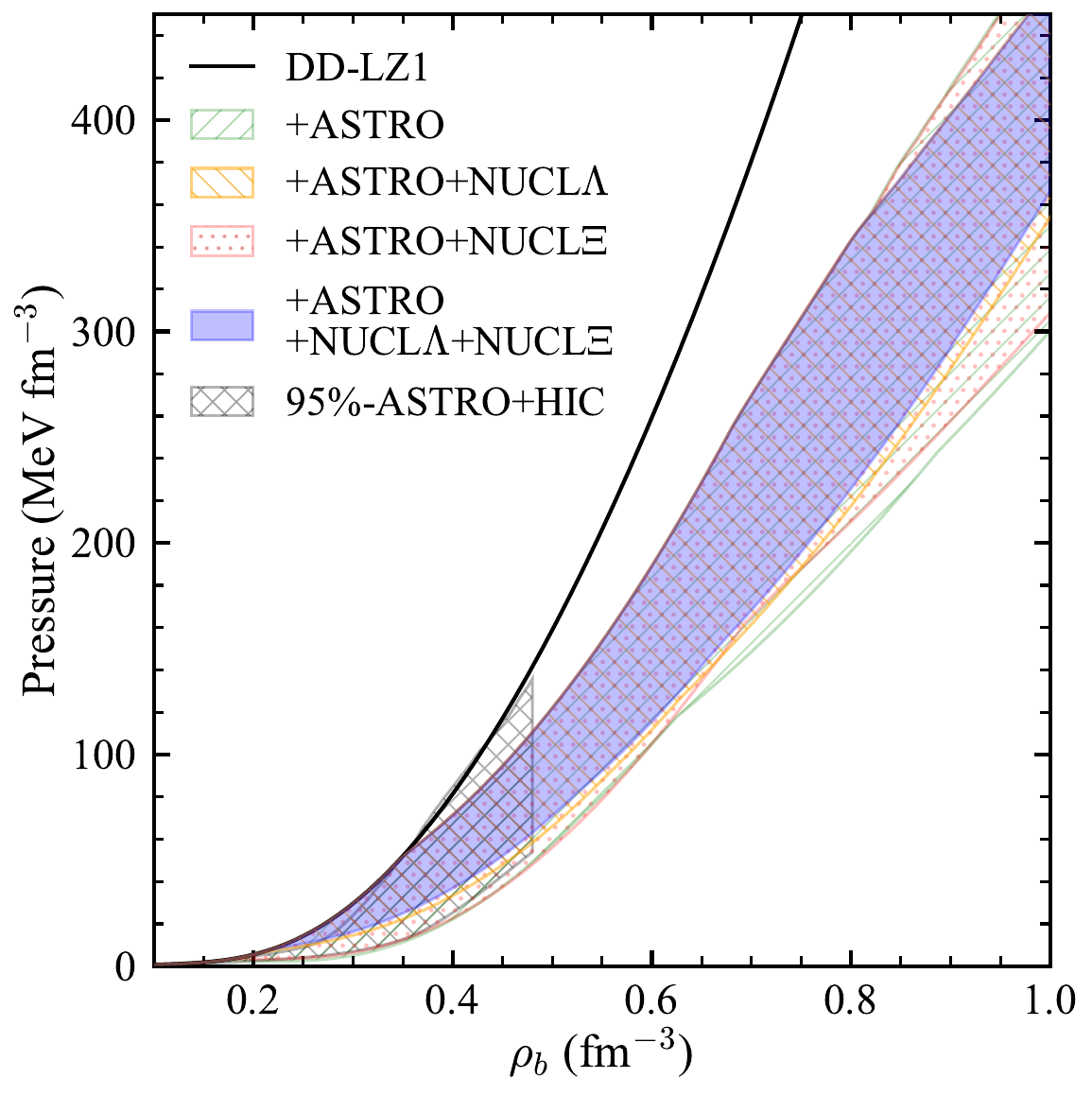}
	\caption{The most probable pressure vs. density ($\rho_{b}$) relations of hyperonic star matter under different constraints are compared with the relation for NS matter (black curve). All results are obtained using the effective interaction DD-LZ1. For reference, the black grid-patterned region in the figure indicates the constraint region derived from combined multimessenger NS observations and heavy-ion collision data \cite{Huth2022Nature606.276}.}\label{Fig:Pressure}
\end{figure}

Under the combined constraints of hypernuclear experiments and astronomical observations, the parameters of $YN$ interactions have been significantly adjusted, directly affecting the EOS of hyperonic star matter. To more clearly illustrate the impact of different astrophysical and nuclear physics constraints, we constructed the pressure-density relations of hyperonic star matter using the representative Lagrangian DD-LZ1 under four different constraint scenarios, based on the coupling strength ratios $R_{\sigma Y}$ and $R_{\omega Y}$ listed in Table \ref{Tab:CouplingStrength}. These results were then compared with those obtained for purely nucleonic matter, as shown in Fig. \ref{Fig:Pressure}. For reference, the empirical constraint region derived from joint analyses of multimessenger NS observations and heavy-ion collision experiments is also marked in the figure (black grid area) \cite{Huth2022Nature606.276}. As shown, tighter constraints lead to a significant reduction in theoretical uncertainties, especially for density regions higher than two times of normal nuclear density. 

\begin{figure}[htbp]
	\centering
	\includegraphics[width=0.48\textwidth]{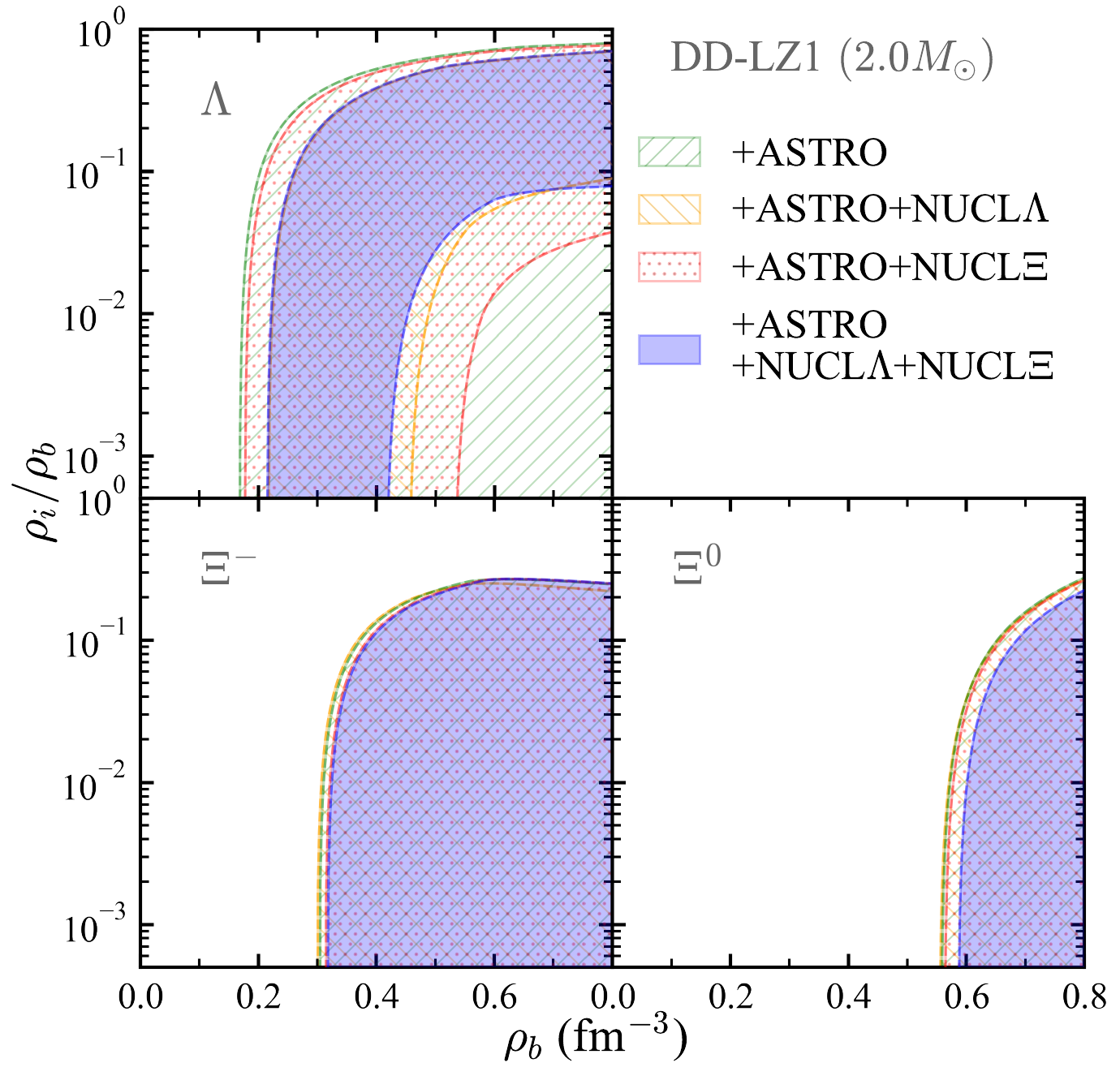}
	\caption{The $\Lambda$ and $\Xi$ hyperon fractions are shown as functions of the baryon density $\rho_{b}$, calculated using the DD-LZ1 model under four different scenarios. Only parameter sets that yield a maximum hyperon star mass of exactly $2M_{\odot}$ are considered.}\label{Fig:FracHyperon}
\end{figure}

Due to significant variations in the EOS under different constraint conditions, these differences inevitably impact the model’s predictions for both the internal composition and macroscopic properties of hyperon stars. Fig. \ref{Fig:FracHyperon} shows the compositions of hyperon stars predicted by all parameter sets yielding a mass of two solar masses configuration, using the stiffest model (DD-LZ1), correspondingly the largest hyperon star parameter prior space, under each test conditions. Since this study primarily focuses on the influence of $\Xi$ hypernuclear constraints on hyperon star properties, only the $\Lambda$ and $\Xi$ hyperons are included in the composition analysis. As shown in the figure, when only astrophysical constraints are considered, the threshold for $\Lambda$ hyperons appears at a baryon density of approximately $\rho_{b}=0.2~\rm{fm}^{-3}$, and increases slightly when hypernuclear constraints are incorporated. Additionally, apart from $\Lambda$ hyperons, the distributions of $\Xi^{-}$ and $\Xi^{0}$ remain largely consistent when hypernuclear physics constraints are included, while the range of $\Lambda$ hyperon distribution gradually narrows with increasing nuclear physics constraints, demonstrating the effectiveness of combined analysis for the determination of key hyperon interaction parameters.

\begin{table*}[hbpt]
	\footnotesize 
	\centering
	\caption{For the four RMF effective interactions employed, the most probable values for the properties of hyperon stars under different constraints are presented with 68\% confidence intervals. For comparison, the corresponding results for pure NSs are also shown. Here, $M_{\rm max}/M_{\odot}$ denotes the maximum mass, $\rho_c$ the corresponding central density, $R_{2.0}$ the radius of a $2.0 M_{\odot}$ star, and $R_{1.4}$ and $\Lambda_{1.4}$ the radius and tidal deformability of a $1.4 M_{\odot}$ star, respectively.}\label{Tab:HyperonStarProperties1}
	\renewcommand{\arraystretch}{1.5}
	\doublerulesep 0.1pt \tabcolsep 9pt
	\begin{tabular}{lcccccc}
		\hline\hline
		&                                & $M_{\rm{max}}/M_{\odot}$  & $\rho_{c}$/fm$^{-3}$      & $R_{2.0}$/km               & $R_{1.4}$/km               & $\Lambda_{1.4}$               \\ \hline
		\multirow{5}{*}{PKDD}   & w.o. $Y$                       & 2.329                     & 0.889                     & 13.223                     & 13.709                     & 765.163                       \\
		& +ASTRO                         & 2.079$^{+0.000}_{-0.203}$ & 0.933$^{+0.058}_{-0.050}$ & 12.762$^{+0.002}_{-0.621}$ & 13.709$^{+0.000}_{-0.000}$ & 765.095$^{+0.000}_{-0.041}$   \\
		& +ASTRO+NUCL$\Lambda$           & 2.055$^{+0.024}_{-0.124}$ & 0.958$^{+0.193}_{-0.025}$ & 12.562$^{+0.202}_{-1.195}$ & 13.709$^{+0.000}_{-0.748}$ & 765.095$^{+0.000}_{-284.469}$ \\
		& +ASTRO+NUCL$\Xi$               & 2.079$^{+0.000}_{-0.167}$ & 0.933$^{+0.065}_{-0.061}$ & 12.762$^{+0.003}_{-0.869}$ & 13.709$^{+0.000}_{-0.061}$ & 765.095$^{+0.000}_{-31.050}$  \\
		& +ASTRO+NUCL$\Lambda$+NUCL$\Xi$ & 2.053$^{+0.026}_{-0.120}$ & 0.962$^{+0.207}_{-0.183}$ & 12.533$^{+0.232}_{-1.225}$ & 13.709$^{+0.000}_{-0.955}$ & 765.095$^{+0.000}_{-340.825}$ \\ \hline
		\multirow{5}{*}{DD-ME2} & w.o. $Y$                       & 2.486                     & 0.817                     & 13.288                     & 13.239                     & 712.009                       \\
		& +ASTRO                         & 2.249$^{+0.000}_{-0.350}$ & 0.873$^{+0.313}_{-0.109}$ & 13.131$^{+0.002}_{-2.682}$ & 13.239$^{+0.000}_{-1.544}$ & 711.874$^{+0.000}_{-434.234}$ \\
		& +ASTRO+NUCL$\Lambda$           & 2.133$^{+0.116}_{-0.209}$ & 0.925$^{+0.251}_{-0.137}$ & 12.882$^{+0.251}_{-1.905}$ & 13.239$^{+0.000}_{-0.832}$ & 711.874$^{+0.000}_{-300.744}$ \\
		& +ASTRO+NUCL$\Xi$               & 2.248$^{+0.001}_{-0.356}$ & 0.873$^{+0.249}_{-0.112}$ & 13.131$^{+0.002}_{-2.080}$ & 13.239$^{+0.000}_{-0.596}$ & 711.874$^{+0.000}_{-231.839}$ \\
		& +ASTRO+NUCL$\Lambda$+NUCL$\Xi$ & 2.154$^{+0.095}_{-0.230}$ & 0.915$^{+0.261}_{-0.142}$ & 12.959$^{+0.174}_{-1.786}$ & 13.239$^{+0.000}_{-0.663}$ & 711.874$^{+0.000}_{-255.074}$ \\ \hline
		\multirow{5}{*}{DD-MEX} & w.o. $Y$                       & 2.554                     & 0.779                     & 13.493                     & 13.341                     & 759.665                       \\
		& +ASTRO                         & 2.303$^{+0.009}_{-0.417}$ & 0.824$^{+0.340}_{-0.091}$ & 13.386$^{+0.001}_{-2.880}$ & 13.340$^{+0.000}_{-2.303}$ & 758.993$^{+0.000}_{-546.485}$ \\
		& +ASTRO+NUCL$\Lambda$           & 2.158$^{+0.154}_{-0.260}$ & 0.894$^{+0.263}_{-0.149}$ & 13.133$^{+0.254}_{-2.151}$ & 13.340$^{+0.000}_{-1.436}$ & 758.993$^{+0.000}_{-453.523}$ \\
		& +ASTRO+NUCL$\Xi$               & 2.288$^{+0.024}_{-0.401}$ & 0.831$^{+0.320}_{-0.124}$ & 13.387$^{+0.000}_{-2.969}$ & 13.340$^{+0.000}_{-2.200}$ & 758.993$^{+0.000}_{-539.662}$ \\
		& +ASTRO+NUCL$\Lambda$+NUCL$\Xi$ & 2.188$^{+0.124}_{-0.310}$ & 0.883$^{+0.219}_{-0.144}$ & 13.211$^{+0.176}_{-2.022}$ & 13.340$^{+0.000}_{-0.797}$ & 758.993$^{+0.000}_{-305.776}$ \\ \hline
		\multirow{5}{*}{DD-LZ1} & w.o. $Y$                       & 2.560                     & 0.779                     & 13.393                     & 13.172                     & 727.908                       \\
		& +ASTRO                         & 2.311$^{+0.005}_{-0.472}$ & 0.830$^{+0.329}_{-0.106}$ & 13.281$^{+0.000}_{-2.944}$ & 13.172$^{+0.000}_{-2.477}$ & 727.698$^{+0.000}_{-538.093}$ \\
		& +ASTRO+NUCL$\Lambda$           & 2.160$^{+0.156}_{-0.255}$ & 0.908$^{+0.213}_{-0.188}$ & 12.993$^{+0.288}_{-2.059}$ & 13.172$^{+0.000}_{-1.458}$ & 727.698$^{+0.000}_{-439.976}$ \\
		& +ASTRO+NUCL$\Xi$               & 2.301$^{+0.015}_{-0.404}$ & 0.836$^{+0.329}_{-0.110}$ & 13.280$^{+0.002}_{-2.857}$ & 13.172$^{+0.000}_{-2.424}$ & 727.698$^{+0.000}_{-532.516}$ \\
		& +ASTRO+NUCL$\Lambda$+NUCL$\Xi$ & 2.178$^{+0.138}_{-0.336}$ & 0.899$^{+0.224}_{-0.143}$ & 13.050$^{+0.231}_{-2.225}$ & 13.172$^{+0.000}_{-1.185}$ & 727.698$^{+0.000}_{-392.640}$ \\ 	\hline\hline
	\end{tabular}
\end{table*}

\begin{figure}[htbp]
	\centering
	\includegraphics[width=0.48\textwidth]{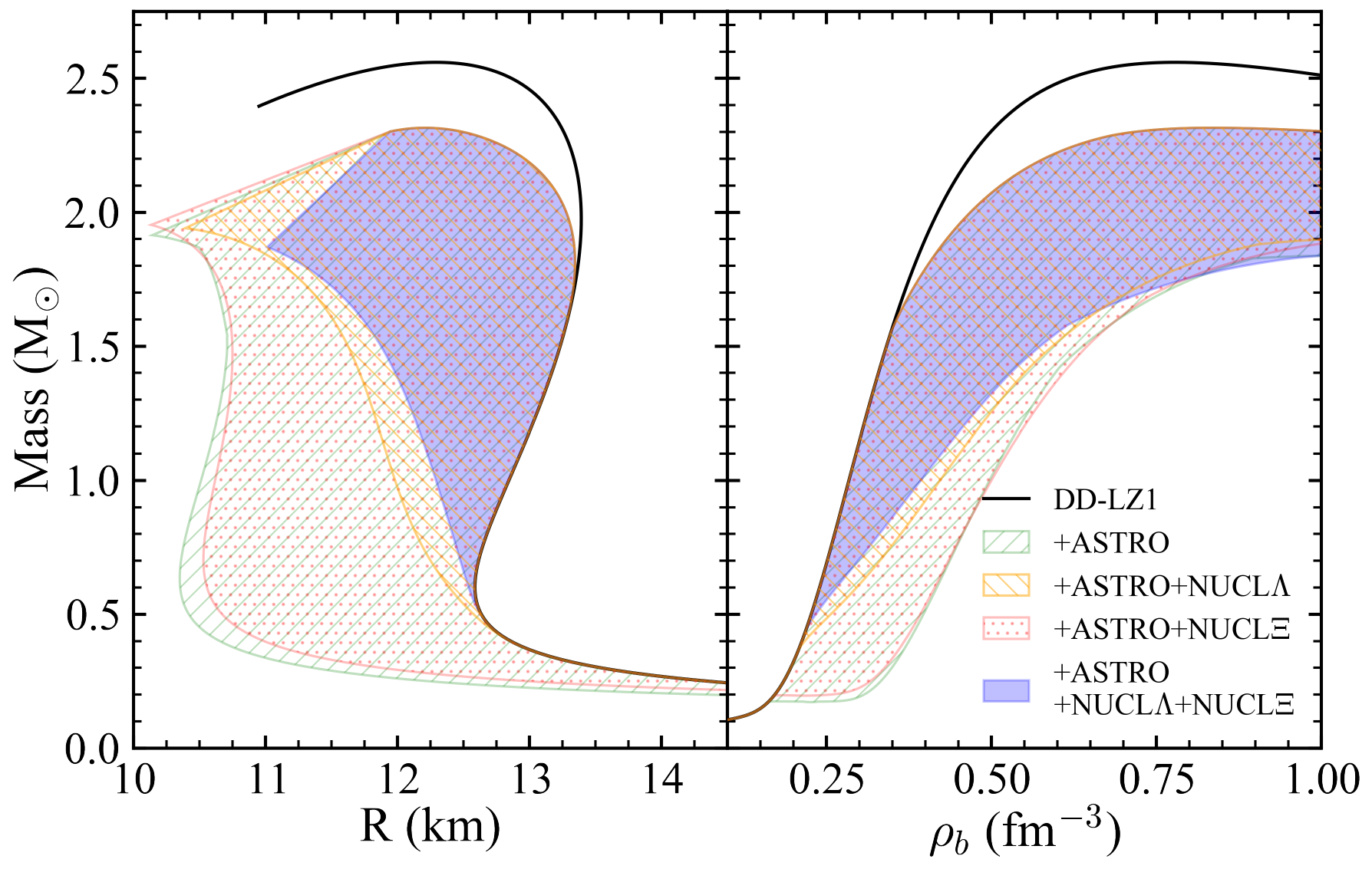}
	\caption{The most probable mass-radius ($R$) and mass-density ($\rho_b$) relations of hyperon stars under different constraints are compared with those of NSs (black curve). All results are obtained using the effective interaction DD-LZ1.}\label{Fig:HyperStarMR}
\end{figure}

Finally, based on the most probable values and 68\% confidence intervals of the hyperon coupling strength ratios $R_{\sigma Y}$ and $R_{\omega Y}$ presented in Table \ref{Tab:CouplingStrength}, we have systematically computed the properties of hyperon stars predicted by various RMF effective Lagrangian models under four testing scenarios (see Section \ref{Sec:test}). The results are summarized in Table \ref{Tab:HyperonStarProperties1}. To illustrate the effects of different constraint conditions on the macroscopic properties of hyperon stars, Fig. \ref{Fig:HyperStarMR} displays the mass-radius and mass-density relations under the four scenarios using the DD-LZ1 model as an example, incorporating the parameter space of hyperon coupling strength ratios from Table \ref{Tab:CouplingStrength}. For comparison, the corresponding predictions for pure NSs are also shown. As expected, the inclusion of hyperons significantly softens the EOS of NS matter, leading to a reduction in the predicted maximum mass, often accompanied by a smaller stellar radius \cite{Long2012PRC85.025806}. When only astrophysical constraints are imposed, the model predictions for the mass-radius and mass-density relations exhibit a wider spread. Upon introducing nuclear physics constraints, the hyperon coupling strengths tend to take larger values, making the emergence of hyperons in NS interiors more difficult and significantly raising the threshold conditions for their appearance \cite{Long2012PRC85.025806}. This change leads to a clear narrowing of the distribution ranges for the mass-radius/density relations, namely, the radius uncertainties of NSs with a certain mass (especially lower mass) are significantly reduced. As the nuclear physics constraints are progressively strengthened, this limiting effect becomes increasingly pronounced, further reducing the uncertainties in the predicted related properties. Ultimately, at the 68\% confidence level, from the joint +ASTRONOMY+NUCL$\Lambda$+NUCL$\Xi$ analysis, the DD-LZ1 model yields a maximum hyperonic star mass around $2.2 M_{\odot}$.

\section{Summary} \label{sec:summary}

Hyperon interactions are pivotal in determining the internal structure and EOS of hyperon stars. While theoretical studies support the presence of hyperons in high-density environments, their interaction mechanisms remain largely uncertain. To address questions such as the description of nuclear systems with strangeness or the hyperon puzzle in NSs, we conduct a statistical study integrating (hyper-)nuclear experiments and NS observations, and the uncertainty of the possible $YN$ contributions can be reliably estimated. This estimate is then compared to the previous calculations that take only the lightest $\Lambda$ hyperons into account Ref. \cite{Sun2023APJ942.55}.

The present systematic work has led to several interesting findings. The study found that when only astronomical observational constraints were considered, the vector coupling strength of hyperons tended to be larger, while the scalar coupling strength leaned toward smaller values. After introducing hypernuclear physics constraints, the parameter space for the scalar and vector coupling strength ratios of both $\Lambda$ and $\Xi$ hyperons contracted significantly. Among these, the adjustment range of the scalar $\sigma$-hyperon coupling was more pronounced, with its coupling strength ratio notably increasing, a trend also supported by calculations based on the structure of $\Xi$ hypernuclei. Further analysis revealed that when constraints from both $\Lambda$ and $\Xi$ hypernuclear physics were considered, due to the shallower potential and larger rest mass of $\Xi$ hyperons, their threshold for appearance inside hyperon stars was higher, leading to a relatively weaker impact on the internal composition and macroscopic properties of the stars. Therefore, compared to $\Lambda$ hyperons, the parameter space for the $YN$ interaction of $\Xi$ hyperons is more loosely constrained. Variations in the $YN$ interaction under different constraints directly affect the model's description of the EOS, thereby affecting predictions about the internal composition, and macroscopic properties of hyperon stars. Based on the stiffest DD-LZ1 model considered, the predicted maximum mass of hyperon stars is $2.178^{+0.138}_{-0.336}~M_{\odot}$ (68\% credible interval).

The results demonstrate that as past and upcoming constraints are incorporated, theoretical uncertainties in predicting the internal structure and macroscopic properties of hyperon stars can be progressively reduced in the future. These findings highlight the critical importance of multi-source data integration in constraining the EOS and hyperonic star models and provide clear guidance for future research. With continuous advancements in astronomical observation techniques and the accumulation of high-precision particle physics data, more accurate and comprehensive experimental inputs are anticipated. These developments are expected to drive significant progress in both astrophysics and hypernuclear physics. Ultimately, they may offer new insights into resolving the "hyperon puzzle" and lay a solid foundation for understanding the behavior of matter under extreme conditions. Through this work, we have demonstrated that the Bayesian approach provides strong constraints not only on the properties of hyperon stars but also on hypernuclear structure properties such as the hyperon potential and separation energy, establishing it as a highly promising method for future research.

Although the current research providing a relatively in-depth exploration of the impact of $\Xi N$ interactions in hyperon stars, the uncertainties in model predictions remain inadequately addressed. Current models of $\Xi N$ interactions largely rely on limited experimental data from identified hypernuclear states, without adequately accounting for the potential uncertainties arising from configuration distributions within $\Xi$ hypernuclei. Furthermore, due to a lack of reliable experimental constraints, $\Sigma$ hyperons \cite{Sedrakian2020PRD102.041301}, as well as hyperon-hyperon interactions, in dense matter are often treated in a simplified manner or entirely neglected, which can further impact the accuracy of theoretical models. Our analysis also does not include potential high-density effects such as $\Delta$ resonances or quark deconfinement, which could become particularly important under extreme conditions. Furthermore, the approach of performing a joint analysis of nucleon-nucleon and hyperon-nucleon interactions also deserves further investigation \cite{Malik2022PRD106.063024, Vishal2025PRD112.023016}. To achieve a more comprehensive understanding of NSs with hyperons, future research must expand theoretical explorations of hyperon interactions, especially through systematic studies across a wider range of densities.

\begin{acknowledgments}
This work was partly supported by the National SKA Program of China (2020SKA0120300), the National Natural Science Foundation of China (11875152, 11873040, 12273028 and 12494572), and the Fundamental Research Funds for the Central Universities, Lanzhou University (lzujbky-2023-stlt01).
\end{acknowledgments}

\end{document}